\documentclass[prd,english,preprintnumbers,amsmath,amssymb,nofootinbib,twocolumn]{revtex4-1}

\pdfoutput=1

\usepackage[latin1]{inputenc}
\usepackage{graphicx}
\usepackage{bbm}
\usepackage{amssymb}
\usepackage{amsmath}

\usepackage{dsfont}

\def\0#1#2{\frac{#1}{#2}}

\def\s0#1#2{\mbox{\small{$ \frac{#1}{#2} $}}}

\newcommand{\Tr}{\mathrm{Tr}}

\newcommand{\E}{\mathrm{e}}

\newcommand{\be}{\begin{eqnarray}}
\newcommand{\ee}{\end{eqnarray}}

\newcommand{\nn}{\nonumber }

\newcommand{\om}{\omega} 

\usepackage{babel}
\makeatother
\begin{document}

\title{Towards a Renormalization Group Approach to Density Functional Theory\\ -- General Formalism and Case Studies --}

\author{Sandra Kemler}\affiliation{Institut f\"ur Kernphysik (Theoriezentrum), Technische Universit\"at Darmstadt, 
D-64289 Darmstadt, Germany}
\author{Jens Braun}\affiliation{Institut f\"ur Kernphysik (Theoriezentrum), Technische Universit\"at Darmstadt, 
D-64289 Darmstadt, Germany}
\affiliation{ExtreMe Matter Institute EMMI, GSI, Planckstra{\ss}e 1, D-64291 Darmstadt, Germany}
  
\begin{abstract}
We discuss a two-point particle irreducible ($2$PPI) approach to many-body physics which relies on 
a renormalization group (RG) flow equation for the associated effective action. In particular, the general structure and properties of this
RG flow equation are analyzed in detail. Moreover, we 
discuss how our $2$PPI RG approach relates to Density Functional Theory and argue that it can in principle be
used to study ground-state properties of non-relativistic many-body systems from microscopic interactions, such as (heavy) nuclei. 
For illustration purposes, we use our formalism to compute the ground-state properties of two toy models.  
\end{abstract}

\maketitle
%
\section{Introduction}
Originally, Density Functional Theory (DFT) has been invented by Hohenberg and 
Kohn to study atoms with many electrons in an efficient way~\cite{Hohenberg:1964zz,Kohn:1965zzb}. 
Since then, DFT has indeed been successfully used to study a large variety of quantum systems with many degrees of freedom, ranging
from electronic systems over ultracold Fermi gases to (heavy) nuclei.

For heavy nuclei, DFT remains currently to be the only feasible approach for a calculation of ground-state properties~\cite{Bender:2003jk}. 
In fact, the nuclear energy density functional approach represents  a very active research field, also documented by the impressive efforts
undertaken by the {UNEDF SciDAC Collaboration~\cite{RJF,Nam:2012gy}.}

The application of DFT to the nuclear many-body problem has been very successful in recent years.
Apart from conceptional advances aiming at, e.~g., ab-initio studies of heavy nuclei, 
nuclear DFT studies provided us with a universal understanding of properties of nuclei which is relevant for a large variety of applications,
see, e.~g., Refs.~\cite{Duguet:2007be,Lesinski:2007ys,Bender:2009ty,Lesinski:2011rn} and also Ref.~\cite{Dobaczewski:2010gr} for a review.

For some time, DFT approaches have been based on fitting the parameters of a given ansatz for the density functional such that one
reproduces the experimentally determined values of the ground-state properties of various heavy nuclei~\cite{Dobaczewski:2001ed}. 
The resulting density functionals have then been employed to describe ground-state properties of other heavy nuclei. 
In recent years, there have been many attempts to give microscopic constraints on the nuclear energy density functional 
employing, e.~g., (improved) density matrix expansions~\cite{Carlsson:2008gm,Stoitsov:2010ha}. The latter approach has been further pursued and used to
derive a (microscopic) nuclear energy density functional from chiral two- and three-nucleon interactions~\cite{Kaiser:2009me,Holt:2011nj}.
{On the other hand, the density-matrix expansion has been tested against ab initio calculations of trapped neutron drops~\cite{Bogner:2011kp}.}
These developments based on density-matrix expansions might be viewed as the beginning of a new era in the context of the nuclear energy density functional approach.
Moreover, the future construction of density functionals will certainly benefit from various different
approaches and complementary studies, ranging from a direct optimization of energy density functionals to
studies of the equation of state of nuclear matter, see, e.~g., Refs.~\cite{Hebeler:2009iv,Hebeler:2010xb,Kortelainen:2010hv,Holt:2011jd,Tews:2012fj}.
At the present stage, however, our understanding of the constraints for energy density functionals emerging from microscopic
nuclear forces as well as the (direct) relation of the energy density functional to these forces
is not yet fully complete and requires further research. An ab-initio renormalization group (RG) approach to DFT could complement and extend these 
efforts~\cite{Schwenk:2004hm,Braun:2011pp} since it may open up the possibility to directly compute ground-state properties of (heavy) nuclei 
from the underlying microscopic nucleon-nucleon interactions in a systematic fashion. In particular, such a functional RG 
approach seems to be promising since it allows to directly study the change  
of the energy density functional under `RG transformations', e.~g., from a weakly-interacting or even non-interacting system (starting point of the RG flow)
to the fully interacting system, namely the nucleus under consideration~\cite{Schwenk:2004hm,Braun:2011pp}. 
Moreover, such 
an RG approach is advantageous as it can be directly related to the underlying path integral for which many systematic approximation
schemes are known, ranging from perturbative schemes to non-perturbative resummation techniques. 
For reviews and introductions to DFT approaches using the path-integral formalism, we refer the reader to Refs.~\cite{Furnstahl:2007xm,Drut:2009ce}. 

From a very field-theoretical point of view, the object of interest is the so-called (quantum) effective action which can be derived from the 
(exponentiated) path integral by introducing source terms and performing then a Legendre transformation with respect to theses sources, see also 
our discussion below. Such a construction is well-known in statistical physics where the Gibbs free energy of, e.g., a spin system, 
is computed from the underlying partition function by means of a Legendre transformation. In quantum field theory, it is possible to work along
these lines and couple the sources to the (physically) relevant ``degrees of freedom". For DFT, this means that we couple the sources to the
densities associated with the quantum fields and then perform a Legendre transformation with respect to these sources. The effective action resulting
form such a procedure is, strictly speaking, called a two-particle point-irreducible (2PPI) effective action.
This 2PPI effective action is directly related to the energy density functional introduced by Hohenberg and Kohn, 
see our discussion below and also Refs.~\cite{Furnstahl:2007xm,Drut:2009ce} for a review.

At this point, we have traced back the computation of the energy density functional to the computation of the path integral of the 
underlying theory defined by the microscopic interactions which itself represents an inherently difficult problem. Due to the relation of the 
energy density functional and the path integral, however, we can utilize powerful existing tools for the computation of the associated path integral. 
For example, we could employ {\it ab-initio} Monte-Carlo (MC) calculations or RG approaches. In the
present work, we consider an RG approach to DFT that has been put forward in Refs.~\cite{Schwenk:2004hm,Braun:2011pp}.
For a more general discussion of the properties of $n$PPI effective actions, we refer the reader to Ref.~\cite{Pawlowski:2005xe}.

In the present work, we give a detailed discussion of the RG approach to DFT introduced in Refs.~\cite{Schwenk:2004hm,Braun:2011pp}. In Sect.~\ref{sec:RGDFT},
we present a general discussion of the structure and properties of these DFT-RG flows,
including their connection to perturbation theory as well as the Hartree and Hartree-Fock approximation.
In Sect.~\ref{sec:cs}, we then apply our DFT-RG approach (2PPI-RG) approach to two simple toy models for which analytic solutions are known. In 
Sect.~\ref{sec:conc}, we finally present our conclusions and outlook, including a concise discussion of our next steps towards 
an application our DFT-RG approach to the nuclear many-body problem.

%
\section{Renormalization Group Approach to Density Functional Theory}\label{sec:RGDFT}
In this section, we give a more general discussion of field-theoretical aspects of the 
DFT-RG approach put forward in Ref.~\cite{Schwenk:2004hm} and further discussed in Ref.~\cite{Braun:2011pp}.
\subsection{RG flow equation}

In nuclear DFT, we are particularly interested in strongly-interacting many-body systems (far) away from the continuum limit
where the ground-state density is inhomogeneous.\footnote{Note that strong interactions are by no means necessary to induce inhomogeneous
ground states. For example, the ground-state density of $N$ non-interacting particles trapped in a harmonic oscillator potential is obviously
inhomogeneous.} More generally speaking, we aim at a study
of a finite system of fermions interacting via a non-local interaction which may be repulsive at short distances and attractive at long range, 
as it is the case for the nuclear many-body problem and ultracold trapped Fermi gases. 
As discussed above, DFT has indeed proven to be useful for studies of these type of systems, see, e.~g., 
Refs.~\cite{Papenbrock:2005bd,Dobaczewski:2010gr,Bulgac:2010dg}.

Before we discuss the DFT-RG flow equation and its properties, we 
give a brief summary of the underlying principles of DFT. To this end, we restrict ourselves to the following 
action:\footnote{Throughout this work, we use the imaginary-time formalism. The extent of the imaginary-time axis can then be
identified with the inverse temperature~$\beta = 1/T$.}
\be
&& S[\psi^{\dagger},\psi] = 
\int_{\tau}\int_x\, \psi^{\dagger}_{\sigma}(\tau,\vec{x}) 
\left[ \partial _{\tau} -  \Delta + V(\vec{x}) \right] \psi_{\sigma}(\tau,\vec{x})\nn\\
&& 
+\,\frac{1}{2} 
\int _{\tau}\!\int _x\! \int _y \,
\psi^{\dagger}_{\sigma}(\tau,\vec{x}) \psi^{\dagger}_{\sigma^{\prime}}(\tau^{\prime}\!,\vec{y}^{\,\prime})
U(\vec{x},\vec{y})
\psi_{\sigma^{\prime}}(\tau^{\prime}\!,\vec{y}^{\,\prime}) \psi_{\sigma}(\tau,\vec{x})\,, \nn 
\label{eq:genclassaction}
\ee
where we have set~$2m=1$ and introduced the shorthands $\int_{\tau}=\int _0 ^{\beta}d\tau$, $\int _{x}=\int d^{d}x$, $d$ is the space dimension.
The function $V(\vec{x})$ denotes a (background) potential.
Moreover, we assume here and in the following that we sum over identical spin indices, if not stated otherwise.
In the following we only consider two-body interactions. Higher-order $N$-body interactions will be ignored but
can be included straightforwardly in our DFT-RG approach.

DFT is based on the famous {\it Hohenberg-Kohn theorem}~\cite{Hohenberg:1964zz}. For a given interaction potential~$U$, 
this theorem states that there exists a one-to-one correspondence between the ground-state density and the potential~$V(\vec{x})$ 
(up to an additive constant), at least for non-degenerate ground states.
This implies that the ground-state density~$n_{\rm gs}$ (uniquely)
determines the ground-state wave-function of the $N$-body problem under consideration. The latter can therefore be considered 
as a functional of the density~$n$. Moreover, the expectation value of any physical observable 
is determined by a unique functional of the ground-state density. In particular, this is true for the ground-state energy of the system and implies the existence
of an energy density functional~$E[n]$. The ground-state density~$n_{\rm gs}(\vec{x})$ 
can then be obtained by minimizing~$E[n]$ with respect to the density:
\be
E_{\rm gs}=\inf_n E[n]\,.
\ee
Moreover, it can be shown that the energy density functional in the limit of vanishing external potential~$V$,
the so-called {\it Hohenberg-Kohn} functional~$E_{\rm HK}$, is {\it universal} for a {\it given} interaction potential~$U$:
\be
E_{\rm HK}[n]=E[n] - \int d^d x\, n(\vec{x})V(\vec{x})\,.
\ee
These considerations can be generalized to the case
of degenerate ground states. However, we leave aside a discussion of the issue of $V$-representability, and also of $N$-representability, 
in the following. We refer the {reader to, e.~g., Refs.~\cite{DreizlerGross,primer,EngelDreizler} for a more detailed discussion.}

The {\it Hohenberg-Kohn} theorem can be viewed as a starting point for an efficient description of many-body problems. However, the theorem
does by no means provide a recipe for the computation of the {\it Hohenberg-Kohn} functional. Similar to the $1$PI quantum effective action in 
conventional quantum field theory, the Hohenberg-Kohn functional consists of infinitely many terms. As we shall see below, the {\it Hohenberg-Kohn} 
functional is indeed closely related to a specific type of effective action.
This observation implies that it is in general not possible to write down the exact {\it Hohenberg-Kohn} functional for a given many-body problem.
Thus, an ansatz for the functional is required in order to determine the ground-state properties of a given many-body problem.
Usually, it is difficult to find a systematic and stable approximation scheme. For example, the simplest approximation is the so-called local
density approximation~(LDA) which can be derived straightforwardly from the density dependence of the ground-state energy of the corresponding
uniform many-body problem. The latter can, for instance, be computed with {\it ab-initio} MC simulations. 
It is indeed possible to show that LDA represents the lowest order in a systematic derivative expansion of the exact
energy-density functional~\cite{Leeuwen}. However, a low-order approximation of this type might only be justified in systems with weakly varying 
densities, such as ultracold Fermi gases with a large number of atoms in an isotropic trap~\cite{Mueller1}. For a general many-body problem, such
a derivative expansion may have bad convergence properties.

Now we would like to make contact between DFT and the effective action approach to quantum field theory, see~, e.~g., Refs.~\cite{Furnstahl:2007xm,Drut:2009ce} for 
a more detailed introduction. The generating functional for our theory defined by the action~$S$ is given by\footnote{Here, we have dropped
an irrelevant normalization factor of the path integral.}
\be
Z [\{J_{\sigma}\}]
&\sim&\!\!\int  \mathcal{D}\psi^{\dagger}  \mathcal{D}\psi\, \E ^{-S[\psi^{\dagger},\psi] + 
\int_{\tau}\int_x J_{\sigma}(\tau,\vec{x})
( \psi^{\dagger}_{\sigma}(\tau,\vec{x}) \psi_{\sigma}(\tau,\vec{x}))
}\nn\\
&\equiv& \E ^{W[\{J_{\sigma}\}]}\,. \label{eq:pathint}
\ee
In contrast to the conventional textbook approach to quantum field theory, namely the 1PI formalism, 
we have coupled the external sources~$\{J _{\sigma}\}$ to terms which are bilinear in the fermion fields.\footnote{Note that 
the formalism can be generalized by including sources coupled to, e.~g., pairing densities. The latter might be considered
as convenient effective degrees of freedom to describe the ground-state properties of certain many-body systems.} 
These bilinears play the role of 
composite bosonic degrees of freedom. 

Let us add a word on the issue of fixing the particle number in a path-integral approach. We can either 
introduce chemical potentials into the path integral to fix the numbers of the various particle species, say protons and neutrons, or
we do not include chemical potentials 
but fix the particle numbers by choosing appropriate boundary conditions for the equations of motion, as discussed in Ref.~\cite{Puglia2003145}. 
In a concrete DFT-RG study, we shall follow the latter approach to fix the particle number~\cite{BSP}.

We now introduce the so-called  {\it classical} fields~$\rho _{\sigma}(\tau,\vec{x})$ which are defined as the (functional) derivative of~$W[\{J_{\sigma}\}]$ with respect to the 
corresponding sources~$J_{\sigma}(\tau,\vec{x})$:
\be
\rho _{\sigma}(\tau,\vec{x}) = \frac{\delta W[\{J_{\sigma}\}]}{\delta J_{\sigma}(\tau,\vec{x})}\,.\label{eq:rhodef}
\ee
Note that~$\rho_{\sigma}$ is not only a function of~$\tau$ and~$\vec{x}$ but also 
a functional of the sources $\{ J_{\sigma}\}$, i.~e.~$\rho_{\sigma}=\rho_{\sigma}[\{ J_{\sigma}\}]$.
Clearly, these fields are related to the particle densities.
In complete analogy to the textbook derivation of the 1PI effective action, the 2PPI effective action is now defined as the
Legendre transformation of~$W$ with respect to the sources~$J_{\sigma}$:
\be
&&\Gamma[\{\rho_{\sigma}\}] \nn\\
&&\;\; = \sup_{\{J_{\sigma}\}} \left\{ -W[\{J_{\sigma}\}] + \int _{\tau}  \int _x\, J_{\sigma}(\tau,\vec{x}) \rho_{\sigma}(\tau,\vec{x}) 
\right\}.
\label{eq:effactdef}
\ee
The so-defined $2$PPI effective action~$\Gamma[\{\rho_{\sigma}\}]$ determines completely the dynamics of the many-body
problem and, up to a factor of~$\beta$, it can be associated with the energy-density functional mentioned above in the context of the conventional 
{\it Hohenberg-Kohn} DFT 
formalism. We add that the exact equivalent of the energy density functional as introduced by Hohenberg and Kohn can be derived similarly, if one introduces
time-independent sources~$J_{\sigma}(\vec{x})$, see, e.~g., Refs.~\cite{PTP.92.833,1997cond.mat..2247V,PhysRevB.66.155113,Puglia2003145}. 
{For a more general discussion on DFT in terms of a Legendre transformation, we refer the reader 
to Refs.~\cite{Eschrig,Kutzelnigg2006163}.}

It is straightforward to show that the $2$PPI effective action~$\Gamma[\{\rho_{\sigma}\}]$ does
not dependent on the sources~$\{J_{\sigma}\}$:
\be
\frac{\delta\Gamma[\{\rho_{\sigma}\}]}{\delta J_{\sigma}}=0\,. \label{eq:statcond}
\ee
On the other hand, we have
\be
\frac{\delta\Gamma[\{\rho_{\sigma}\}]}{\delta \rho_{\sigma} (\tau,\vec{x})}=J_{\sigma}(\tau,\vec{x})\,.\label{eq:gsfunc}
\ee
Thus, the ground-state configuration~$\{\rho_{\sigma,\rm gs}\}$ 
is determined by this equation in the limit~$J_{\sigma}\to 0$.\footnote{In the limit of vanishing sources, Eq.~\eqref{eq:gsfunc} represents
the quantum equation of motion of the composite degrees of freedom~$\rho_{\sigma}$.}
In other words, solving Eq.~\eqref{eq:gsfunc} for the fields $\rho_{\sigma}(\tau,\vec{x})$ in this limit, 
we find the ground-state configuration~$\{\rho_{\sigma,\rm gs}\}$. We define the (time-independent) ground-state densities~$n_{\rm gs,\sigma}(\vec{x})$
as follows: 
\be
{n}_{\rm gs,\sigma}(\vec{x}) := \frac{1}{\beta}\int _0^{\beta}d\tau\, \rho_{\rm gs,\sigma}(\tau,\vec{x})\,.
\ee
If the solutions~$\{\rho_{\sigma}(\tau,\vec{x})\}$ turn out to be independent of the imaginary time~$\tau$, then
we have ${n}_{\sigma,\rm gs}(\vec{x}) \equiv \rho_{\sigma,\rm gs}(\tau,\vec{x})$. In our toy model studies to be discussed below, this
is indeed the case.\footnote{In fact, this $\tau$-independence of~$\rho_{\rm gs}$ follows from our flow
equation~\eqref{eq:DFTfloweq} to be discussed below, provided that we only consider an interaction potential~$U$ with 
a imaginary-time dependence of the form~$U\sim\delta(\tau-\tau^{\prime})$.}

From the solutions~$\{\rho_{\sigma}(\tau,\vec{x})\}$, we eventually obtain the ground-state energy~$E_{\rm gs}$ of
the system:
\be
E_{\rm gs}:=\lim_{\beta\to\infty}\frac{1}{\beta}\Gamma[\{\rho_{\sigma,\rm gs}\}]\,. \label{eq:egsgen}
\ee
This relation follows immediately from the spectral representation of the partition function~$Z$: $Z\sim \sum_n {\rm e}^{-\beta E_n}$,
and~$\Gamma \sim -W[J]|_{\{J_{\sigma}\to 0\}} \sim \ln Z[J]|_{\{J_{\sigma}\to 0\}}$. From Eq.~\eqref{eq:egsgen}, we can also anticipate
the relation between the effective action~$\Gamma$ and the {\it Hohenberg-Kohn} functional~$E_{\rm HK}$:
$\Gamma \sim \beta E_{\rm HK}$.

At this point, we would like to add that the {\it universality} of the {\it Hohenberg-Kohn} functional~$E_{\rm HK}$ 
follows from the fact that background potential can be absorbed
into the source terms~$J_{\sigma}$ by a simple shift, $J_{\sigma}\to J_{\sigma}+V$, see Ref.~\cite{Schwenk:2004hm}. 
Exploiting this observation, we find
\be
&&\!\!\!\!\!\Gamma[\{\rho_{\sigma}\}] \nn \\
&&=\Gamma_{\rm HK}[\{\rho_{\sigma}\}]\!+\! \sum_{\sigma}\int _0 ^{\beta} \! d\tau \int\! d^dx\, V(\vec{x})\rho_{\sigma}(\tau,\vec{x})\,,
\ee
where~$\Gamma_{\rm HK}[\{\rho_{\sigma}\}]=\Gamma_{V=0}[\{\rho_{\sigma}\}]$. 
We conclude that the functional~$\Gamma_{\rm HK}[\{\rho_{\sigma}\}]$ depends only
on our choice for the interaction potential but not on the background potential~$V$. Recall that in our case the~$\rho_{\sigma}$'s depend in
general on the imaginary time~$\tau$, in contrast to the standard {\it Hohenberg-Kohn} functional which depends only on a time-independent
density. In this respect, our present argument can be viewed as a trivial generalization of the original universality argument given
by {Hohenberg} and {Kohn}.

As discussed above, the computation of the effective action~$\Gamma$ ($\sim$ {\it Hohenberg-Kohn} functional) 
for a given theory can be an inherently difficult task. In the present work, we employ a DFT-RG approach that has been put forward in
Refs.~\cite{PhysRevB.66.155113,Schwenk:2004hm,Braun:2011pp}. For details on the derivation, we also refer the reader to these papers. 
For convenience, we only consider a system of $N$~spinless fermions in the following. However, we stress that the derivation of the flow equation
is by no means bound to such a theory but can be straightforwardly generalized to other non-relativistic theories. 
To be more specific, we consider a classical action of the following general form: 
\begin{widetext}
\be
&& S_{\lambda}[\psi^{\ast},\psi] = \int_{\tau} \int _x \, \psi^{\ast}(\tau,\vec{x}) 
\left[ \partial _{\tau} - \Delta +V_{\lambda}(\vec{x}) \right] \psi (\tau,\vec{x})
+\,\frac{\lambda}{2} \int _{\tau} \int _x \int _{\tau^{\prime}} \int _{x^{\prime}} \,
\psi^{\ast}(\tau,\vec{x})  \psi^{\ast}(\tau^{\prime},\vec{x}^{\,\prime}) 
U(\tau,\tau^{\prime};\vec{x},\vec{x}^{\prime})
\psi (\tau^{\prime},\vec{x}^{\,\prime})\psi(\tau,\vec{x})
\,\label{eq:action}\nn
\ee
\end{widetext}
with $2m\equiv 1$.
Here, we allow for a very general form of the two-body interaction. In most cases, however,
one will restrict~$U$ to be of the form~$U\sim \delta(\tau-\tau^{\prime})\tilde{U}(\vec{x},\vec{x}^{\,\prime})$.
The parameter~$\lambda \in [0,1]$ denotes a dimensionless control parameter: For~$\lambda=0$, 
the two-body interaction potential~$U$
is turned off and we are left with an exactly soluble non-interacting problem, namely $N$ fermions trapped in 
a given potential~$V_{\lambda}$. For~$\lambda=1$, the potential~$U$ is fully turned on and the potential~$V_{\lambda}$ 
has assumed its physical form. For example, $V_{\lambda}$ can be chosen such that it plays the role of 
the trap potential in an experiment with ultracold Fermi gases. In studies of ground-state properties of
 self-bound many-body problems, such as nuclei, one chooses~$V_{\lambda=1}(\vec{x})\equiv 0$.
Apart from the physical constraint at~$\lambda=1$, the form of~$V_{\lambda}$ is at our disposal and can be chosen
such that, e.~g., the initial non-interacting problem is simple to solve. For example, we could choose~$V_{\lambda=0}$ to be a harmonic potential.

We add that, from a field-theoretical point of view, the one-body potential~$V_{\lambda}$ acts as a regulator function. 
In fact, the length scale associated with the potential~$V_{\lambda}$ sets a momentum
scale in the theory which removes infrared divergences in the appearing loop diagrams. In this sense, the introduction of the parameter~$\lambda$
renders the theory scale-dependent.
In particular for the case~$V_{\lambda=1}(\vec{x})\equiv 0$, we shall also assume that the Fourier transform of the 
interaction potential~$U$ falls off sufficiently rapidly for large momenta to avoid the occurrence of ultraviolet divergences. 

Following Refs.~\cite{Schwenk:2004hm,Braun:2011pp}, it is now straightforward to
derive the RG flow equation for the `scale-dependent' $2$PPI effective action~$\Gamma_{\lambda}[\rho]$. 
By taking the derivative of~$\Gamma_{\lambda}[\rho]$
with respect to~$\lambda$, we find 
\be
\partial_{\lambda} \Gamma_{\lambda}[\rho] &=&(\partial_{\lambda} V_{\lambda}) \cdot \rho
+\frac{1}{2}  \rho\cdot U \cdot \rho \nn\\
&& \quad\qquad + \frac{1}{2}\,{\rm Tr}\, U\cdot \left( \frac{\delta^2 \Gamma_{\lambda}[\rho]}{\delta \rho\delta \rho}\right)^{-1},
\label{eq:DFTfloweq}
\ee
where the dot represents a shorthand for
\be
A\cdot B\equiv \int _0^{\beta}d\tau\int d^dx A(\tau,\vec{x})B(\tau,\vec{x})\,
\ee
and the trace $\rm Tr$ stands for
\be
{\rm Tr}\, M(\tau^{\prime},\vec{x}^{\,\prime},\tau,\vec{x}) = \int _0^{\beta}d\tau\int d^dx M(\tau,\vec{x},\tau,\vec{x})\,.
\ee
This functional differential equation~\eqref{eq:DFTfloweq} describes the flow from the non-interacting system defined at $\lambda=0$ to the
interacting theory defined at the physical point~$\lambda=1$.  The effective action~$\Gamma_{\lambda=0}[\rho]$ associated with 
an exactly soluble (non-interacting) $N$-body problem determines the initial condition of the RG flow. 

In the terminology of many-body physics, the second term in Eq.~\eqref{eq:DFTfloweq} can be identified as the so-called Hartree term. The third term on the right-hand side
depends on the scale-dependent density-density correlator $\delta^2\Gamma_{\lambda}[\rho]/(\delta \rho\delta \rho)$ 
and includes all other corrections to the 
effective action, also so-called Fock contributions, see also our discussion in Sect.~\ref{sec:hartree}.

We observe that the flow equation~\eqref{eq:DFTfloweq} has a simple one-loop structure as it is the case for the RG flow
equation for the $1$PI effective action derived by Wetterich~\cite{Wetterich:1992yh,Berges:2000ew}. However, this does not mean that we can only capture
one-loop corrections with this flow equation. On the contrary, by solving the functional differential equation~\eqref{eq:DFTfloweq}, we automatically
include corrections of arbitrarily high-order, see also our discussion in Sect.~\ref{sec:pt}. 
In fact, no approximations are involved in the derivation of the flow equation~\eqref{eq:DFTfloweq}. 
In particular, the derivation of this flow equation does not require that the interaction strength is small. 
The equation is exact, provided we only allow for two-body interactions in the classical
action~$S_{\lambda}$.\footnote{A generalization of this flow equation to include the effects of 
higher $n$-body operators is straightforward. In principle, this requires to also multiply the $3-,\,4-,\dots$ body-interaction terms
with the control parameter~$\lambda$.}

For a general many-body problem, 
it is not possible to solve the flow equation~\eqref{eq:DFTfloweq} for the $2$PPI effective action (corresponding to the {\it Hohenberg-Kohn} functional) 
exactly. Thus, we need to find (systematic) approximation/truncation schemes. 
Examples for such approximation schemes are the gradient expansion of the effective action or an 
expansion of~$\Gamma_{\lambda}[\rho]$ about the ground state~$\rho_{\rm gs,\lambda}$: 
\begin{widetext}
\be
\!\!\!\Gamma_{\lambda}[\rho]=\Gamma_{\lambda}[\rho_{\rm gs,\lambda}] 
+ \frac{1}{2}\int _{\tau}\int _x \int_{\tau^{\prime}}\int_{x^{\prime}} \left( \rho(\tau,\vec{x}) 
- \rho_{\rm gs,\lambda}(\tau,\vec{x})\right)\Gamma^{(2)}_{\lambda}[\rho_{\rm gs,\lambda}](\tau,\vec{x};\tau^{\prime},\vec{x}^{\,\prime}) \left( \rho(\tau^{\prime},\vec{x}^{\,\prime}) 
- \rho_{\rm gs,\lambda}(\tau^{\prime},\vec{x}^{\,\prime})\right)+\dots\,. \label{eq:vertexexp}
\ee
\end{widetext}
Plugging this expansion into the general flow equation~\eqref{eq:DFTfloweq}, we obtain the flow equations 
for~$\Gamma_{\lambda}[\rho_{\rm gs,\lambda}]$, $\rho_{\rm gs,\lambda}$, and~$\Gamma^{(n)}_{\lambda}[\rho_{\rm gs,\lambda}]$ with~$n\geq 2$.
We then find that the flow of~$\rho_{{\rm gs},\lambda}$ depends on~$\rho_{{\rm gs},\lambda}$ itself but also on~$\Gamma^{(2)}_{\lambda}[\rho_{\rm gs,\lambda}]$
and~$\Gamma^{(3)}_{\lambda}[\rho_{\rm gs,\lambda}]$. The flow equation for the two-point function~$\Gamma^{(2)}_{\lambda}[\rho_{\rm gs,\lambda}]$ depends
on~$\rho_{{\rm gs},\lambda}$, $\Gamma^{(2)}_{\lambda}[\rho_{\rm gs,\lambda}]$, $\Gamma^{(3)}_{\lambda}[\rho_{\rm gs,\lambda}]$, 
and $\Gamma^{(4)}_{\lambda}[\rho_{\rm gs,\lambda}]$. In general,
we find that the flow of~$\Gamma^{(n)}_{\lambda}[\rho_{\rm gs,\lambda}]$ depends on~$\rho_{{\rm gs},\lambda}$ 
and~$\Gamma^{(m)}_{\lambda}[\rho_{\rm gs,\lambda}]$ with~\mbox{$m=2,\dots,n,n+1,n+2$}.

In the terminology of quantum field theory, 
the expansion~\eqref{eq:vertexexp} corresponds to a vertex expansion and will be discussed in detail in our toy model studies in Sect.~\ref{sec:cs}. 
Loosely speaking, the $n$-point correlation function is related to the expectation value of $n$ density operators:
\be
&&\Gamma^{(n)}_{\lambda}[\rho_{\rm gs,\lambda}](\tau_{1},\vec{x}_{1};\dots;\tau_{n},\vec{x}_{n})\nn\\
&&\qquad\qquad \sim 
\langle 
\hat{\rho}(\tau_1,\vec{x}_1)\hat{\rho}(\tau_2,\vec{x}_2)\cdots \hat{\rho}(\tau_n,\vec{x}_n)
\rangle_{\rm gs}\,,
\label{eq:gnexpval}
\ee
where
\be
\hat{\rho} (\tau,\vec{x})=\psi^{\ast}(\tau,\vec{x})\psi(\tau,\vec{x})\,.\label{eq:densop}
\ee
For brevity, we have dropped the subtraction of the 
disconnected contributions on the right-hand side of Eq.~\eqref{eq:gnexpval}, see 
Eq.~\eqref{eq:propdensop} for a more rigorous expression for~$n=2$.
At this point, we would like to stress that the vertex expansion given in Eq.~\eqref{eq:vertexexp}
is an exact expansion of the functional~$\Gamma_{\lambda}[\rho]$ 
about the ground state and should by no means be confused with the local
density approximation. Recall that the expansion coefficients, namely the $n$-point functions, are
functions of the time-like and spatial coordinates.

Before we discuss the features of our RG flow equation, we would like to emphasize that our present approach allows us to 
directly compute the energy density functional from the microscopic interactions of the theory. To be concise, it opens up the possibility
to derive the energy density functional (and the ground-state properties) of, say, heavy nuclei from chiral effective field theory 
interactions~\cite{Epelbaum:2002ji,Epelbaum:2002vt,Epelbaum:2005pn,Navratil:2007we,Epelbaum:2008ga} as the latter can be used to determine
the interaction potential~$U$ in our flow equation.

\subsection{Excited States}
Up to this point, we have only discussed how to compute the ground-state energy. However, it is
also possible to extract the energy of the excited states of the theory from the $2$PPI effective 
action~$\Gamma[\{\rho_{\sigma}\}]\equiv \Gamma_{\lambda=1}[\{\rho_{\sigma}\}]$. To this end,
it comes to our rescue that we consider the system at a finite temperature~$T$. Using now the spectral representation of the 
partition function~$Z$, we find
\be
\Gamma[\{\rho_{\sigma,\rm gs}\}] = - \ln \left( \sum_{n}{\rm e}^{-\beta E_n} \right)+{\rm const.}\,.\label{eq:specrep}
\ee
In order to obtain the energy of the excited states $E_n$ ($n>0$, $E_{\rm gs}\equiv E_0$), we could simply compute
$\Gamma[\{\rho_{\sigma,\rm gs}\}] $ as a function of $\beta$ and then fit the result to the functional form
given on the right-hand side of Eq.~\eqref{eq:specrep} with the energies of the excited states~$E_n$ 
as the free fit parameters. For sufficiently low temperatures (i.~e. sufficiently large~$\beta$), it is reasonable to assume that 
only the lowest-lying
states are occupied. In a fit, we would then only include a small finite number of fit parameters corresponding to these
states. However, such a procedure may still turn out to be impractical from a numerical point of view. 

In order to extract the energy of the first excited state, we do not require a fit procedure involving the spectral representation
of the partition function. In fact, the first excited state~$E_1$ can be extracted from the two-density correlation 
function evaluated at the ground state,
\be
\left.\Gamma^{(2)}_{\sigma,\sigma^{\prime}}(\tau,\vec{x};\tau^{\prime},\vec{x}^{\,\prime}) \right|_{\{\rho_{\sigma,\rm gs}\}}
=\left.\frac{\delta^2 \Gamma[\{\rho_{\sigma}\}]}{\delta\rho_{\sigma}(\tau,\vec{x})\delta\rho_{\sigma^{\prime}}(\tau^{\prime},\vec{x}^{\,\prime})}\right|_{\{\rho_{\sigma,\rm gs}\}}\,.\nn
\ee
First, we note that~$\Gamma^{(2)}_{\sigma,\sigma^{\prime}}$ is directly related to the (field-dependent)
propagator~$G_{\sigma,\sigma^{\prime}}$:
\be
G_{\sigma,\sigma^{\prime}}(\tau,\vec{x};\tau^{\prime},\vec{x}^{\,\prime}) 
=\left (\Gamma^{(2)}_{\sigma,\sigma^{\prime}}(\tau,\vec{x};\tau^{\prime},\vec{x}^{\,\prime}) \right)^{-1}\,.
\ee
On the other hand, the propagator can be written in terms of expectation values of density operators:
\be
&&\!\!\! G_{\sigma,\sigma^{\prime}}(\tau,\vec{x};\tau^{\prime},\vec{x}^{\,\prime})\nn\\
&& \; =\! \left\langle \hat{\rho}_{\sigma}(\tau,\vec{x})\hat{\rho}_{\sigma^{\prime}}(\tau^{\prime},\vec{x}^{\,\prime}) \right\rangle
\!-\! \left\langle \hat{\rho}_{\sigma}(\tau,\vec{x}) \right\rangle   \left\langle \hat{\rho}_{\sigma^{\prime}}(\tau^{\prime},\vec{x}^{\,\prime}) \right\rangle.
\label{eq:propdensop}
\ee
The quantity~$\hat{\rho}_{\sigma}$ is the straightforward generalization of~$\hat{\rho}$ defined in Eq.~\eqref{eq:densop}.
From this expression, it follows that
\be
(E_1-E_{\rm gs})=-\lim_{\beta\to\infty}
\frac{1}{\beta}\ln G_{\sigma,\sigma}(0,\vec{x};\beta,\vec{x}^{\,\prime})\,,
\ee
independent of our choice for the space-time coordinate-pairs~$(\tau,\vec{x})$ and~$(\tau^{\prime},\vec{x}^{\,\prime})$.
Thus, we can extract the energy of the first excited state from the knowledge of the two-point correlation function once we have
computed the ground-state energy using Eq.~\eqref{eq:egsgen}. In principle, it is also possible to project on higher-lying excited
states by considering the expectation values of properly chosen operators. In any case, our discussion shows that
excited states are also accessible within our $2$PPI effective action approach.

\subsection{Perturbation Theory}\label{sec:pt}
Let us now discuss the properties of our DFT-RG approach in more detail. For convenience, we now
restrict ourselves again to the case of spinless fermions, i.~e. we shall consider the flow equation~\eqref{eq:DFTfloweq}
in the following. However, our line of arguments also holds for any type of fields: fermions with spin or even scalar fields
as we shall see in our toy model studies in Sect.~\ref{sec:cs}.

In order to recover perturbation theory from the flow equation~\eqref{eq:DFTfloweq}, we first introduce a counting parameter~$u_0$
into the theory as follows:
\be
U(\tau,\tau^{\prime};\vec{x},\vec{x}^{\,\prime})=u_0 \tilde{U}(\tau,\tau^{\prime};\vec{x},\vec{x}^{\,\prime})\,,
\ee
where
\be
 \tilde{U}(\tau,\tau^{\prime};\vec{x},\vec{x}^{\,\prime})= \phi_{\rm U}(\vec{x},\vec{x}^{\,\prime})\delta (\tau-\tau^{\prime})\,.
\ee
Here, the full dependence on the space coordinates has been absorbed into the function~$\phi_{\rm U}$. For convenience, we assume
that the dependence on the imaginary time is given by~$\delta(\tau-\tau^{\prime})$ and that~$\phi_{\rm U}$ is a dimensionless function.
Now we can introduce a dimensionless counting parameter~$\bar{u}_0$,
\be
\bar{u}_0=\ell_{V}^{2} u_0\,. \label{eq:dimcouplgen}
\ee
Here, $\ell _V$ denotes a typical length scale associated with the background potential~$V$. For example, we have 
$\ell_V=1/\sqrt{\omega}$, if we choose a harmonic background potential: $V(\vec{x})=(1/2)\omega^2\vec{x}^{\,2}$.
Using the length scale~$\ell_V$, we can also introduce the dimensionless ground-state energy, density, as well as 
dimensionless correlation functions~$\Gamma^{(n)}_{\lambda}$:
\be
\bar{E}_{\rm gs}=\ell _{V}^2 E_{\rm gs}\,,\qquad
\bar{\rho}_{\rm gs} = \ell _{V}^{d} \rho_{\rm gs}\,,
\ee
and
\be
\bar{\Gamma}^{(n)} &=& \ell_V^{2n} \Gamma^{(n)}\,,
\ee
where~$\Gamma^{(n)}=\delta^{n}\Gamma/\delta \rho^n$. Note that $\ell_V^{-2}\beta$ is dimensionless.

The dimensionless ground-state energy, density, and correlation functions can be expanded in powers of~$\bar{u}_0$:
\be
\bar{E}_{\rm gs} &=&\sum_{n=0}^{\infty} \epsilon_n \bar{u}_0^n\,, \label{eq:egsexp}\\
\bar{\rho}_{\rm gs} &=& \sum_{n=0}^{\infty} \nu_n \bar{u}_0^n\,,\\
\bar{\Gamma}^{(m)} &=& \sum_{n=0}^{\infty} \gamma^{(m)}_n \bar{u}_0^n \label{eq:gexp}\,.
\ee
The coefficients~$\epsilon_0$, $\nu_0$, and~$\gamma^{(m)}_0$ are determined by the initial condition of the 
flow equation, i.~e. by the functional~$\Gamma_{\lambda=0}$ of the non-interacting theory. At~$\lambda=0$, 
all the expansion coefficients with~$n>0$ are identical to zero.
However, all the coefficients~$\epsilon_n$, $\nu_n$, and~$\gamma^{(m)}_n$ with $n>0$ depend implicitly on~$\lambda$ and may 
therefore be generated dynamically by quantum corrections.
Note that
the coefficients~$\nu_n$ also depend on~$\tau$ and~$\vec{x}$. Analogously, the coefficients~$\gamma^{(m)}_n$ depend
on $m$ pairs $(\tau,\vec{x})$ of space-time coordinates.

Let us now analyze the perturbative expansion of the ground-state energy. To this end, we apply the expansion~\eqref{eq:vertexexp} about
the current ground-state to our general flow equation~\eqref{eq:DFTfloweq}. This yields the following equation for~$\Gamma_{\lambda}[\rho_{\rm gs}]$:
\be
\partial_{\lambda} \Gamma_{\lambda}[\rho_{\rm gs,\lambda}] &=&(\partial_{\lambda} V_{\lambda}) \cdot \rho_{\rm gs,\lambda}
+\frac{1}{2}  \rho_{\rm gs,\lambda} \cdot U \cdot \rho_{\rm gs,\lambda} \nn\\
&& \quad\; +\, \frac{1}{2}\,{\rm Tr}\, U\cdot \left( \left.\frac{\delta^2 \Gamma_{\lambda}[\rho]}{\delta \rho\delta \rho}\right|_{\rho_{\rm gs,\lambda}}\right)^{-1}\!.
\label{eq:floweqgsgen}
\ee
Plugging the expansions~\eqref{eq:egsexp}-\eqref{eq:gexp} into this flow equation and
noting that $E_{\rm gs,\lambda}=\Gamma[\rho_{\rm gs,\lambda}]/\beta$ for $\beta\to\infty$, we can derive flow equations for the coefficients~$\epsilon_n$ by simply 
comparing the left-hand side and right-hand side order by order in our expansion in powers of~$\bar{u}_0$.

Let us now distinguish between two cases, namely the case with a $\lambda$-dependent background potential and the one with
a $\lambda$-independent background potential. We begin our analysis with the latter case. The first term on the right-hand side of 
Eq.~\eqref{eq:floweqgsgen} then vanishes identically.
To the flow of the coefficient~$\epsilon_0$, no term on the right-hand side of Eq.~\eqref{eq:floweqgsgen} contributes
since they depend explicitly on the interaction potential. Thus, these terms can only contribute to the flow of the coefficients~$\epsilon_n$ with $n>0$.

In order to obtain the correct result for the ground-state energy $E_{\rm gs}$ ($\sim \Gamma[\rho_{\rm gs}]$) in leading order in~$\bar{u}_0$,
we deduce from Eq.~\eqref{eq:DFTfloweq} that we need to compute the coefficient~$\gamma^{(2)}_0$
which is associated with the term independent of~$\bar{u}_0$ in the expansion~\eqref{eq:gexp}:
\be
{\rm Tr}\, U\cdot \left( \left.\frac{\delta^2 \Gamma_{\lambda}[\rho]}{\delta \rho\delta \rho}\right|_{\rho_{\rm gs,\lambda}}\right)^{-1} 
\!\!\sim\; \bar{u}_0\, {\rm Tr}\, \tilde{U}\cdot \left(\gamma^{(2)}\right)^{-1}\! + {\mathcal O}(\bar{u}_0^2)\,.\nn
\ee
Interaction-induced corrections of the propagator do not contribute to the leading-order correction of the ground-state energy.
This follows immediately from the fact that the third term on the right-hand side of~Eq.~\eqref{eq:floweqgsgen} depends explicitly on the interaction potential~$U$ and
therefore on~$\bar{u}_0$. Since the second term on the right-hand side has an explicit $U$-dependence as well, the coefficient~$\nu_0$ is also
required to recover the result from perturbation theory at leading order:
\be
 \rho_{\rm gs,\lambda} \cdot U \cdot \rho_{\rm gs,\lambda} \;\sim\; \bar{u}_0 \left(\nu_0 \cdot  \tilde{U} \cdot \nu_0 \right) + {\mathcal O}(\bar{u}_0^2)\,. \nn
\ee

These considerations can be continued successively. In general, we find that we need to compute~$\rho_{\rm gs}$ and the
two-point function up to order~$\bar{u}_0^{n-1}$, in order to obtain the correct result for the ground-state
energy up to order~$\bar{u}_0^n$. 

Let us now turn to the case with a $\lambda$-dependent background potential. We can follow the same line of arguments
as in the previous case. Due to the presence of the 
term~$(\partial_{\lambda}V_{\lambda})\cdot \rho_{\rm gs,\lambda}$, however, we now find that we need to compute~$\rho_{\rm gs,\lambda}$ and the
two-point function up to order~$\bar{u}_0^{n}$ in order to obtain the correct result for the ground-state
energy up to order~$\bar{u}_0^n$.\footnote{In fact, this case is more subtle. Whereas it is apparent that we have to compute~$\rho_{\rm gs}$
up to order~$\bar{u}_0^n$  in order to obtain the correct result for the ground-state energy to the same order, one might naively expect
that we only need to compute the two-point function up to order~$\bar{u}_0^{n-1}$, as in the previous case. However, the computation
of~$\rho_{\rm gs}$ up to order~$\bar{u}_0^n$ requires that we know the two-point function up to order~$\bar{u}_0^{n}$, in the case of
a $\lambda$-dependent background potential. Overall, we therefore need to compute~$\rho_{\rm gs}$ and the
two-point function up to order~$\bar{u}_0^{n}$ in order to obtain the correct result for the ground-state
energy up to this order.}

In summary, we have seen that the computation of the ground-state energy up to a given order~$\bar{u}_0$ requires that we also
compute~$\rho_{\rm gs}$ and the two-point function up to a certain order. The required expansion order for the latter 
depends on whether the background potential is~$\lambda$-dependent or not.

As already mentioned above, the flow equations for the density and the two-point function (and therewith for 
their expansion coefficients) can be obtained by expanding the general flow equation~\eqref{eq:DFTfloweq} about the current ground state~$\rho_{\rm gs,\lambda}$
and then projecting it on the corresponding quantities, namely~$\Gamma_{\lambda}[\rho_{\rm gs,\lambda}]$, $\rho_{\rm gs,\lambda}$, and the $n$-point
correlation functions~$\Gamma_{\lambda}^{(n)}[\rho_{\rm gs,\lambda}]$, see also Eq.~\eqref{eq:vertexexp}. We shall discuss this procedure in great detail 
in our toy model studies in Sect.~\ref{sec:cs}. At this point, however, we already would like to emphasize that the vertex expansion should by no means be
confused with the perturbative series expansion discussed above. The associated expansion coefficients are inherently non-perturbative 
quantities. Below, we use the vertex expansion about the current ground-state since
it is systematic and allows us to extract the RG equations for~$\Gamma_{\lambda}[\rho_{\rm gs,\lambda}]$, $\rho_{\rm gs,\lambda}$, and the $n$-point
correlation functions~$\Gamma_{\lambda}^{(n)}[\rho_{\rm gs,\lambda}]$ in a simple manner.

\subsection{Hatree approximation}\label{sec:hartree}

The so-called Hartree approximation can be obtained from the general flow equation~\eqref{eq:DFTfloweq}
by dropping the third term on the right-hand side. The latter includes, e.~g., the so-called Fock term. Dropping the
third term on the right-hand side of Eq.~\eqref{eq:DFTfloweq}, we can solve the flow equation for~$\Gamma_{\lambda}[\rho]$
analytically. We find
\be
\Gamma_{\lambda}[\rho]=(V_{\lambda}-V_{\lambda=0})\cdot\rho + \frac{\lambda}{2}\rho\cdot U\cdot \rho + \Gamma_{\lambda=0}[\rho]\,,
\label{eq:eahartree}
\ee
where~$\Gamma_{\lambda=0}[\rho]$ is simply the $2$PPI effective action of the non-interacting (initial) system at~$\lambda=0$.
From the definition of the ground-state,
\be
\left. \frac{\delta \Gamma_{\lambda}[\rho]}{\delta \rho} \right|_{\rho_{\rm gs,\lambda}} =0\,,
\ee
we obtain the implicit {equation
\be
\rho_{\rm gs,\lambda}=-\frac{1}{\lambda}\,U^{-1}\cdot  \left [\Delta V_{\lambda}+
\left. \frac{\delta \Gamma_{\lambda=0}[\rho]}{\delta \rho}\right|_{\rho_{\rm gs,\lambda}} \right]\,
\ee
with}~$\Delta V_{\lambda}=(V_{\lambda}\! - \! V_{\lambda=0})$.
Expanding the initial effective action~$\Gamma_{\lambda=0}[\rho]$ about the initial ground-state~$\rho_{\rm gs,0}\equiv\rho_{\rm gs,\lambda=0}$ only up
to second order, we find the following {solution for~$\rho_{\rm gs,\lambda}$:
\be
\!\!\!\!\!\rho_{\rm gs,\lambda}=-\left[{\lambda}U + \Gamma^{(2)}_{{\rm gs},0}\right]^{-1}\!\!\!\!\!\cdot
\left[  \Delta V_{\lambda} \!-\! \Gamma^{(2)}_{{\rm gs},0}\cdot \rho_{\rm gs,0} \right]\,,
\ee
where}~$\Gamma^{(2)}_{{\rm gs},0}:=\Gamma^{(2)}_{{\rm gs},\lambda=0}[\rho_{{\rm gs},\lambda=0}]$. However, such a low-order expasion
in~$\rho$ can only be meaningful if the interacting ground-state~$\rho_{{\rm gs},\lambda=1}$ is close to the initial non-interacting
ground-state~$\rho_{{\rm gs},\lambda=0}$. In general, it is difficult to judge {\it a priori} whether this is the case.\footnote{Strictly speaking, 
the notion `close' requires the definition of a measure on the space defined by the functions~$\rho$. We shall skip this issue here.}
From a practical point
of view, we therefore have to include higher orders in the expansion of~$\Gamma_{\lambda=0}[\rho]$ 
and analyze the convergence of the physical observables as a function of the expansion order.

From our discussion, it follows that the Hartree approximation already yields arbitrarily high orders in an expansion in 
powers of~$\bar{u}_0$, both for the ground-state energy and density. Recall that~$U\sim \bar{u}_0$ and~$E_{\rm gs}\sim \Gamma[\rho_{\rm gs}]$.
However, we would like to point out a shortcoming of the Hartree approximation which becomes apparent from our analysis. 
Taking into account our findings from Sect.~\ref{sec:pt}, we conclude that the Hartree approximation necessarily fails to reproduce the perturbative result for the
ground-state energy, even at leading order. This is simply due to the fact that the third term on the right-hand side of Eq.~\eqref{eq:DFTfloweq} is
missing in this approximation. This term depends explicitly on the interaction potential. 
Since~$\delta^2 \Gamma_{\lambda}/(\delta\rho\delta\rho)$ is in general not identical to zero, even in the non-interacting
limit, this term generates terms which already contribute to the leading order in a perturbative expansion.

\section{Case Studies}\label{sec:cs}
In this work, we refrain from an explicit study of ground-state properties of nuclei but rather present toy model studies 
to explain the theoretical formalism detailed in the previous section. The (classical) actions~$S$ underlying
our toy model studies resemble the action underlying non-relativistic many-body problems in a few aspects, at least from
a purely field-theoretical point of view. However, we also would like to emphasize that such toy model studies only represent
a first step. Studies of simple self-bound many-body models, 
which are also much closer to nuclear physics from a phenomenological point of view,
will be presented elsewhere~\cite{BSP}.

\subsection{Zero-dimensional Toy Model}
The simplest example for the application of our RG approach is the computation of `ordinary' integrals which, loosely speaking, corresponds to zero-dimensional
field theory. In this case, the partition function corresponding to Eq.~\eqref{eq:pathint} is an `ordinary' integral of the form\footnote{Here and
in the following, we drop again irrelevant normalization factors of the partition function. Moreover, we note that the
quantities~$Z[J]$,~$\Gamma[\rho]$, $\dots$
are no functionals but `ordinary' functions in the present case. Nevertheless, we stick to our notation introduced in the previous section.}
\be
Z[J]\sim\int_{-\infty}^{\infty} d\psi\, {\rm e}^{-S[\psi] + J\psi^2}\,,
\ee
where
\be
S[\psi]=\frac{1}{2}\omega^2\psi^2 + \frac{u_0}{24}\psi^4\,. 
\ee
Here,~$J$ and~$\psi$ are real-valued numbers rather than fields. 
Thus, the derivative terms in the action vanish identically. 
Moreover, we have chosen~$U=u_0/12$ for the interaction potential 
and~$V_{\lambda}=(1/2)\omega^2$ for the background potential. We add
that zero-dimensional models have been already successfully employed to benchmark other
field-theoretical methods, see, e.~g., Ref.~\cite{Keitel:2011pn}.
\begin{figure}[t]
\includegraphics[width=\linewidth]{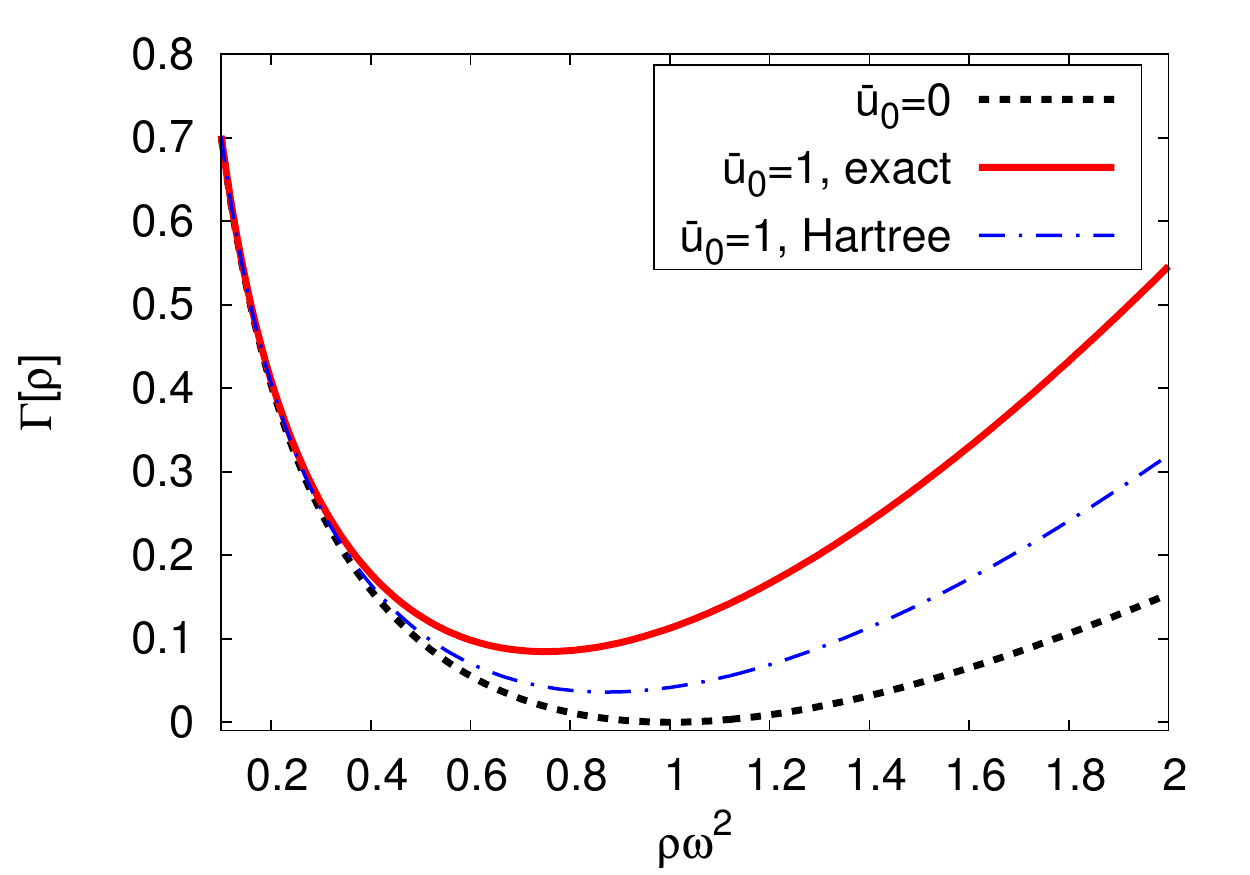}
\caption{Effective action~$\Gamma[\rho]$ of the zero-dimensional toy model for the non-interacting case~$\bar{u}_0=0$ and
for~$\bar{u}_0=1$, as obtained from a direct calculation of the partition function. We have normalized the effective action
such that~\mbox{$\Gamma[\rho_{\rm gs}]=0$} 
for~$\bar{u}_0=0$.}
\label{fig:d0effact}
\end{figure}
\begin{figure*}[t]
\includegraphics[width=0.49\linewidth]{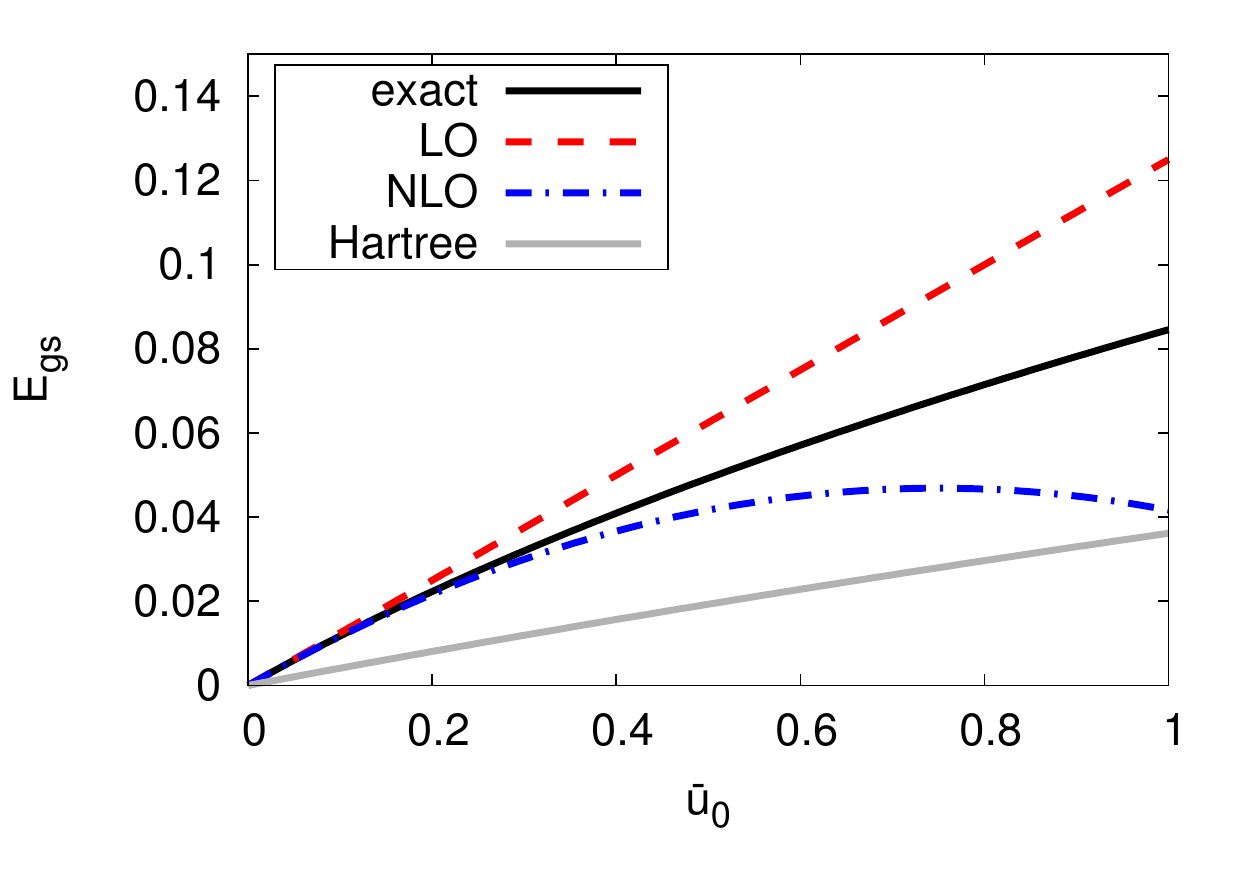}
\includegraphics[width=0.49\linewidth]{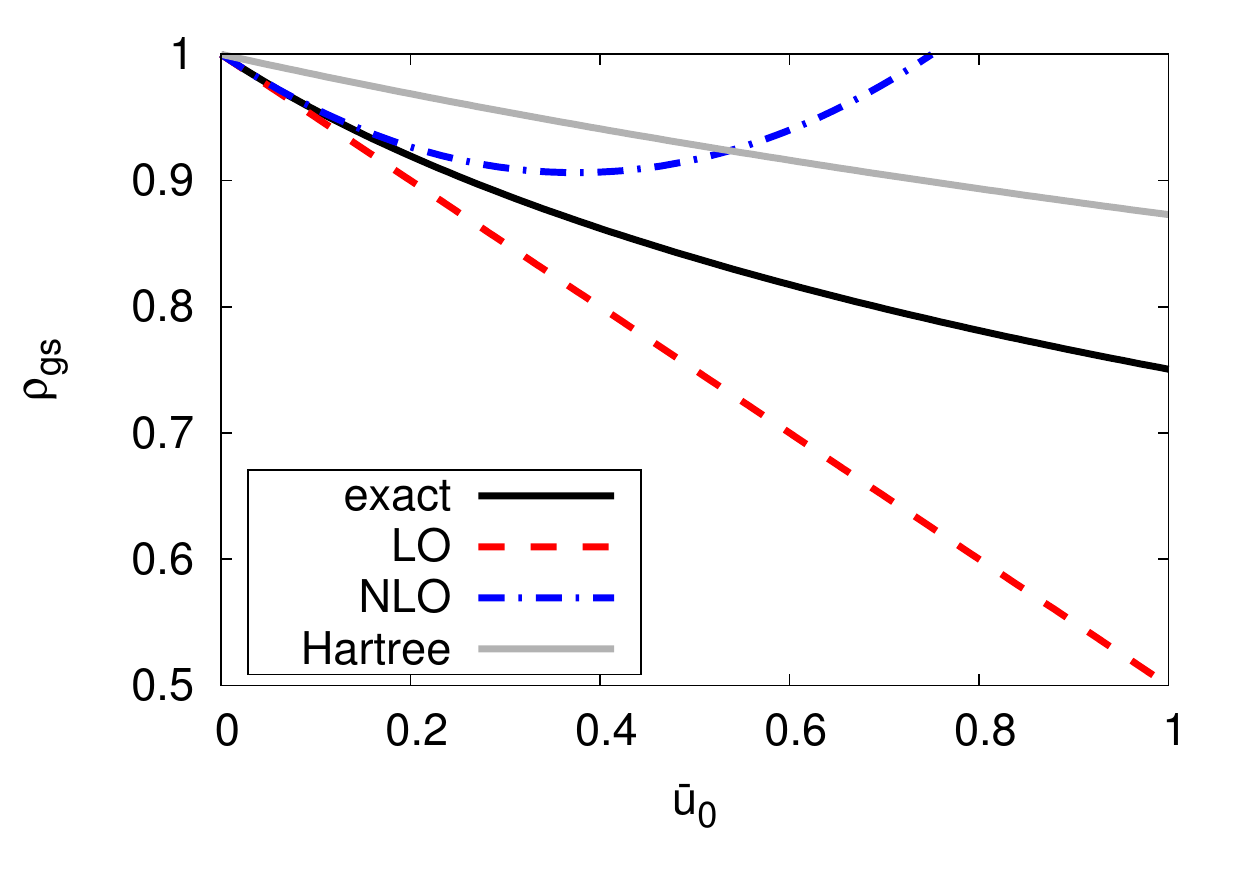}
\caption{Ground-state `energy'~$E_{\rm gs}$ and ground-state `density'~$\rho_{\rm gs}$ of the zero-dimensional toy model
as a function of the (dimensionless) coupling~$\bar{u}_0$. For comparison, we also show the results from  a small coupling
expansion at leading order (LO) and next-to-leading order (NLO).}
\label{fig:d0egs}
\end{figure*}

For~$\bar{J}=J/\omega^2 \leq -1/2$, the partition function can be given in closed form:
\be
\!\!\!\!\! Z[\bar{J}]\!\sim\!
\frac{  
\sqrt{3\!-\! 6 \bar{J} }\,K_{\frac{1}{4}}\left(\frac{3
   (1\!-\! 2 \bar{J})^2}{4 \bar{u}_0}\right){\rm e}^{\frac{3 (1\!-\! 2 \bar{J})^2}{4 \bar{u_0}}}
   }{\sqrt{\bar{u}_0}}
   \equiv {\rm e}^{W[J]},
   \label{eq:zjd0}
\ee   
where $\bar{u}_0=u_0/\omega^4$ and~$K_{\nu}$ is the modified Bessel function of the second kind of order~$\nu$,
see, e.~g., Ref.~\cite{GS}.
For~$\bar{J} > -1/2$, the integral~$Z[J]$ can still be computed numerically.

Using Eq.~\eqref{eq:rhodef} and taking the limit~$J\to 0$, we can now compute  
the ground-state~$\rho_{\rm gs}\equiv\langle \psi^2\rangle_{\rm gs}$:
\be
\rho_{\rm gs}& =&
\left( { \omega^2\bar{u}_0 K_{\frac{1}{4}}\left(\frac{3}{4 \bar{u}_0}\right)}\right)^{-1}
\left( (\bar{u}_0+3)
   K_{\frac{1}{4}}\left(\frac{3}{4 \bar{u}_0}\right) \right.\nn\\
   && 
   \qquad \left. - \frac{3}{2} K_{\frac{5}{4}}\left(\frac{3}{4
   \bar{u}_0}\right)\!-\! \frac{3}{2} K_{-\frac{3}{4}}\left(\frac{3}{4 \bar{u}_0}\right)\right)\,.
\label{eq:rhogsd0}
\ee
This expression can be expanded in powers of~$\bar{u}_0$:
\be
\rho_{\rm gs}
=\omega^{-2}\left(1-\frac{\bar{u}_0}{2}+\frac{2 \bar{u}_0^2}{3}-\frac{11 \bar{u}_0^3}{8}+{\mathcal O}\left(\bar{u}_0^4\right)\right)\,.
\label{eq:rgsd0u0}
\ee
The ground-state `energy' can be obtained directly from
\be
E_{\rm gs}=- \left(\ln Z[0] - \ln Z[0]\big|_{{u}_0\to 0}\right)\,,\label{eq:egsd0}
\ee
where we have normalized~$E_{\rm gs}$ such that it is zero for~$\bar{u}_0\to 0$. For small~$\bar{u}_0$,
we find
\be
E_{\rm gs} = \frac{\bar{u}_0}{8}-\frac{\bar{u}_0^2}{12}+\frac{11 \bar{u}_0^3}{96}+{\mathcal O}\left(\bar{u}_0^4\right)\,.
\ee

Alternatively, we may compute the effective action~$\Gamma[\rho]$ using Eq.~\eqref{eq:effactdef}. From the minimization
of the effective action, we then obtain~$\rho_{\rm gs}$ and~$E_{\rm gs}$. Note that 
the computation of the effective action requires that we solve~$\rho\equiv \rho[J]$, as defined by Eq.~\eqref{eq:rhodef}, for the source~$J$.
The solution~$J=J[\rho]$ needs to be plugged into the definition~\eqref{eq:effactdef} of the effective action.
For the case~$\bar{u}_0=0$, for instance, the
effective action can be computed analytically. We find
\be
\Gamma_{\bar{u}_0=0}[\rho]=\frac{1}{2} \left(\rho\omega^2 -\ln \left({2 \pi  \rho\omega^2}\right)-1\right).
\ee
Apparently, any global analytic ansatz for~$\Gamma_{\bar{u}_0=0}$ is bound to fail. However,
a Taylor expansion about the ground-state~$\rho_{\rm gs}\omega^2=1$ (for $\bar{u}_0=0$) is possible and meaningful.

For~$\bar{u}_0>0$, the computation of the effective action can only be performed numerically. We find that the non-analyticity 
at~$\rho=0$ persists in this case. In Fig.~\ref{fig:d0effact} we show the effective action~$\Gamma[\rho]$ as a function of~$\rho$
for~\mbox{$\bar{u}_0=1$}. We have normalized~$\Gamma[\rho]$ such that~$\Gamma[\rho_{\rm gs}]=0$ for~\mbox{$\bar{u}_0=0$}.
For~$\bar{u}_0=1$, the ground state~$\rho_{\rm gs}$ is found at~$\rho_{\rm gs}= 0.750\dots$, in agreement with our analytic 
result~\eqref{eq:rhogsd0}. For the ground-state `energy', we find~$E_{\rm gs} = 0.084\dots$, which also agrees with
the analytically found value as obtained from Eqs.~\eqref{eq:zjd0} and~\eqref{eq:egsd0}. We can compare these results
with those from the effective action in the Hartree approximation as derived from Eq.~\eqref{eq:eahartree}:
\be
\Gamma_{\rm Hartree}[\rho]=\frac{\bar{u}_0}{24}(\rho\omega^2)^2 + \Gamma_{\bar{u}_0=0}[\rho]+\frac{1}{2}\ln(2\pi)
\,.
\ee
The normalization has been again chosen such that~$\Gamma_{\rm Hartree}[\rho_{\rm gs, Hartree}]=0$ for~$\bar{u}_0\to 0$.
For the ground-state, we find
\be
\rho_{\rm gs, Hartree}(\bar{u}_0)&=&\bar{u}_0^{-1} \left({ \sqrt{6 \bar{u}_0+9}-3}\right)\nn\\
&=& 1-\frac{\bar{u}_0}{6}+\frac{\bar{u}_0^2}{18}-\frac{5 \bar{u}_0^3}{216}+{\mathcal O}\left(\bar{u}_0^4\right)
\,.
\ee
The ground-state energy is then given by
\be
E_{\rm gs}(\bar{u}_0) &=& \Gamma_{\rm Hartree}[\rho_{\rm gs,Hartree}] \nn\\
&=& \frac{\bar{u}_0}{24}-\frac{\bar{u}_0^2}{144}+\frac{5 \bar{u}_0^3}{2592}+{\mathcal O}\left(\bar{u}_0^4\right)\,.
\ee
Clearly, the Hartree approximation does not reproduce the exact results, even in the small-coupling limit. In fact, it
significantly underestimates the exact results for~$E_{\rm gs}$ and overestimates those for 
the ground-state~$\rho_{\rm gs}$.

In Fig.~\ref{fig:d0egs} we show our results for~$E_{\rm gs}$ and~$\rho_{\rm gs}$. 
The comparison of the exact results with the results from a small-coupling expansion at leading and next-to-leading
order shows that the latter are only meaningful for~\mbox{$\bar{u}_0\lesssim 0.2$}. The results from the
Hartree approximation are not in agreement with the exact results, neither at small coupling nor at strong coupling.
For~$\bar{u}_0=1$, for example, the relative error of the Hartree expansion amounts to about~$57\%$.

For illustration purposes, we now compute~$E_{\rm gs}$ and~$\rho_{\rm gs}$ with our flow equation~\eqref{eq:DFTfloweq}.
In the present case, it assumes a simple form:
\be
\partial_{\lambda}\Gamma_{\lambda}[\rho]
=\frac{1}{24}\bar{u}_0\omega^4\left[
 \rho^2 + \left(\frac{\delta \Gamma_{\lambda}[\rho]}{\delta\rho\delta\rho}\right)^{-1}\right]\,.\label{eq:floweqd0}
\ee
In order to solve this equation, we expand~$\Gamma[\rho]$ about the current ground-state~$\rho_{\rm gs,\lambda}$, 
see also Eq.~\eqref{eq:vertexexp}:
\be
&&\!\!\!\Gamma_{\lambda}[\rho]=\Gamma_{\lambda}[\rho_{\rm gs,\lambda}]
+\!\sum_{n=2}^{N_{\rm max}} \frac{1}{n!} \Gamma_{\lambda}^{(n)}[\rho_{\rm gs,\lambda}]
(\rho \!-\!\rho_{\rm gs,\lambda})^n\,,
\label{eq:vertexd0model}
\ee
where~$N_{\rm max}$ denotes the order of the truncation. Note that~$\partial_{\lambda}\Gamma^{(1)}_{\lambda}[\rho_{\rm gs,\lambda}]\equiv 0$ 
and~$\Gamma^{(1)}_{\lambda=0}[\rho_{\rm gs,\lambda}]\equiv 0$ by definition. Plugging this expansion
into the flow equation~\eqref{eq:floweqd0} then yields a tower of flow equations for~$E_{\rm gs,\lambda}=\Gamma_{\lambda}[\rho_{\rm gs,\lambda}]$, $\rho_{\rm gs,\lambda}$, 
and the $n$-point functions~$\Gamma_{\lambda}^{(n)}[\rho_{\rm gs,\lambda}]$. The initial conditions for these
flow equations can be extracted by expanding~$\Gamma_{\bar{u}_0=0}[\rho]$ about its ground state~$\rho_{\rm gs}$. 
\begin{figure}[t]
\includegraphics[width=\linewidth]{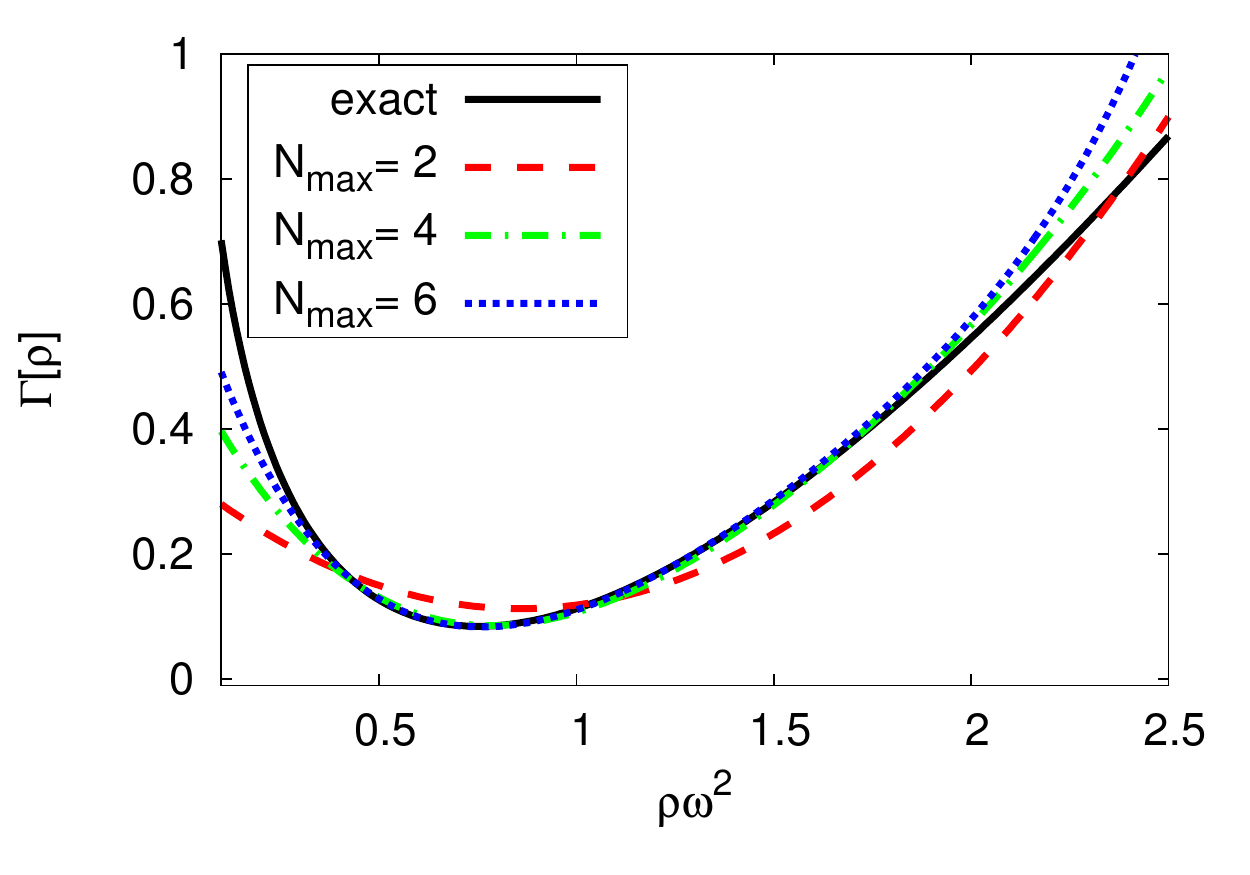}
\caption{Effective action~$\Gamma[\rho]\equiv \Gamma_{\lambda=1}[\rho]$ of the zero-dimensional toy model 
as obtained from our RG approach for various values of~$N_{\rm max}$ for fixed~$\bar{u}_0=1$. 
Again, we have normalized the effective action such that~$\Gamma_{\lambda=1}[\rho_{\rm gs}]=0$ 
for~$\bar{u}_0=0$.}
\label{fig:ead0_rg}
\end{figure}

For~$N_{\max}=2$,
for example, we find the following set of coupled ordinary first-order differential equations:
\be
\partial_{\lambda} E_{\rm gs,\lambda}&=&\frac{1}{24}\bar{u}_0\omega^4 \left[ \rho_{\rm gs}^2 + \left( \Gamma^{(2)}_{\lambda}[\rho_{\rm gs,\lambda}]\right)^{-1}\right]\,,\\
\partial_\lambda\rho_{\rm gs,\lambda} &=& -\frac{1}{12}\bar{u}_0\omega^4 \rho_{\rm gs}  \left( \Gamma^{(2)}_{\lambda}[\rho_{\rm gs,\lambda}]\right)^{-1}\,,\\
\partial_\lambda\Gamma^{(2)}_{\lambda}[\rho_{\rm gs,\lambda}] &=& \frac{1}{12}\bar{u}_0\omega^4\,.
\ee
Note that~$N_{\max}=2$ is sufficient to exactly reproduce the perturbative results for the ground-state energy~$E_{\rm gs}$ 
at leading order. In order to correctly reproduce the leading order of the perturbative expansion of
the ground state~$\rho_{\rm gs}$ and the $n$-point functions of higher order, we need to increase the truncation order beyond~\mbox{$N_{\rm max}=2$}.
From our general discussion above, it follows that $N_{\rm max}=4$ is sufficient to correctly reproduce the perturbative series 
of~$\rho_{\rm gs}$ and~$\Gamma^{(2)}_{\lambda}[\rho_{\rm gs,\lambda}]$ at leading order.
Moreover, we also recover the correct results for~$E_{\rm gs}$
at next-to-leading order with~$N_{\rm max}=4$.

We would like to add that the set of flow equations for~$N_{\rm max}=2$ can be solved analytically. We refrain from giving
the explicit result here. We only state that the ground-state energy behaves as
\be
E_{\rm gs}(\bar{u}_0) \sim \ln \bar{u}_0\,.
\ee
This is in accordance with the asymptotic behavior of the exact result, only that the coefficient of this term is not
reproduced correctly for~$N_{\rm max}=2$.
\begin{figure}[t]
\includegraphics[width=\linewidth]{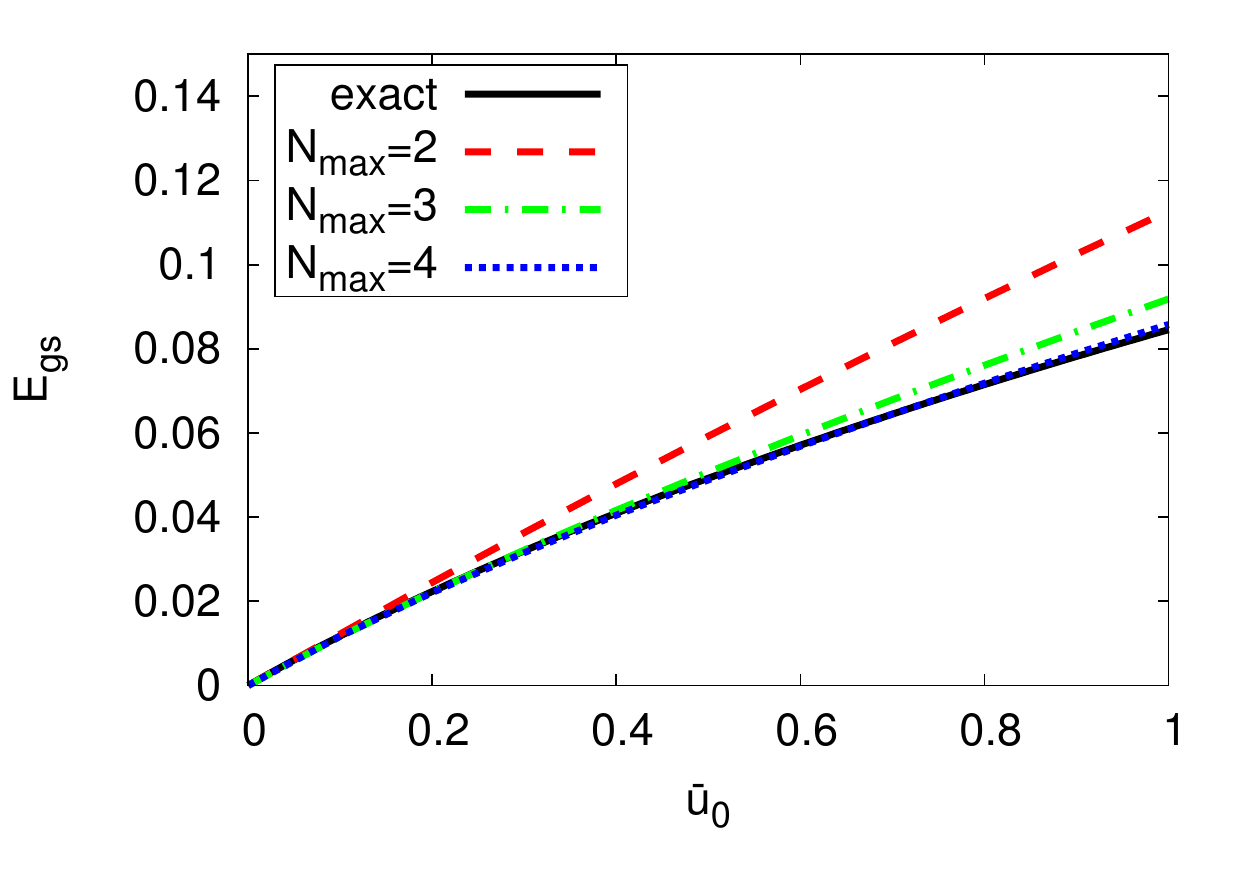}
\caption{$E_{\rm gs}$ as a function of~$\bar{u}_0$ as obtained from our RG study. Note that our results for~$N_{\rm max}=4$
are already almost indistinguishable from the exact values for~$E_{\rm gs}$ on the scale of the plot.}
\label{fig:egsd0_rg}
\end{figure}

In Fig.~\ref{fig:ead0_rg}, we present our results for~$\Gamma[\rho]\equiv \Gamma_{\lambda=1}[\rho]$ as a function of~$\rho$ 
for various values of~$N_{\rm max}$ for fixed~\mbox{$\bar{u}_0=1$}. Note that the radius of convergence
of our expansion~\eqref{eq:vertexd0model} about the ground state is finite. In fact, we find~\mbox{$r_{\rho}/\omega^2=1$} 
for the (dimensionless) radius of convergence in the case~$\bar{u}_0=0$. Our 
numerical results for finite~$\bar{u}_0$ are in accordance with this result. In fact, we do not observe
a convergent behavior around~$\rho\omega^2 \approx 2$ for increasing~$N_{\rm max}$. On the other hand, our results for~$\Gamma[\rho]$
nicely approach the exact results for \mbox{$|\rho - \rho_{\rm gs}|\omega^2 \lesssim 1$} when~$N_{\max}$ is increased.
In order to compute~$\Gamma[\rho]$ for~$\rho\omega^2 > 2$, we could employ
Taylor expansions around various different points with overlapping regions of convergence. This would 
be of importance, for example, when we expect that~$\Gamma[\rho]$ develops various minima.

\begin{figure}[t]
\includegraphics[width=\linewidth]{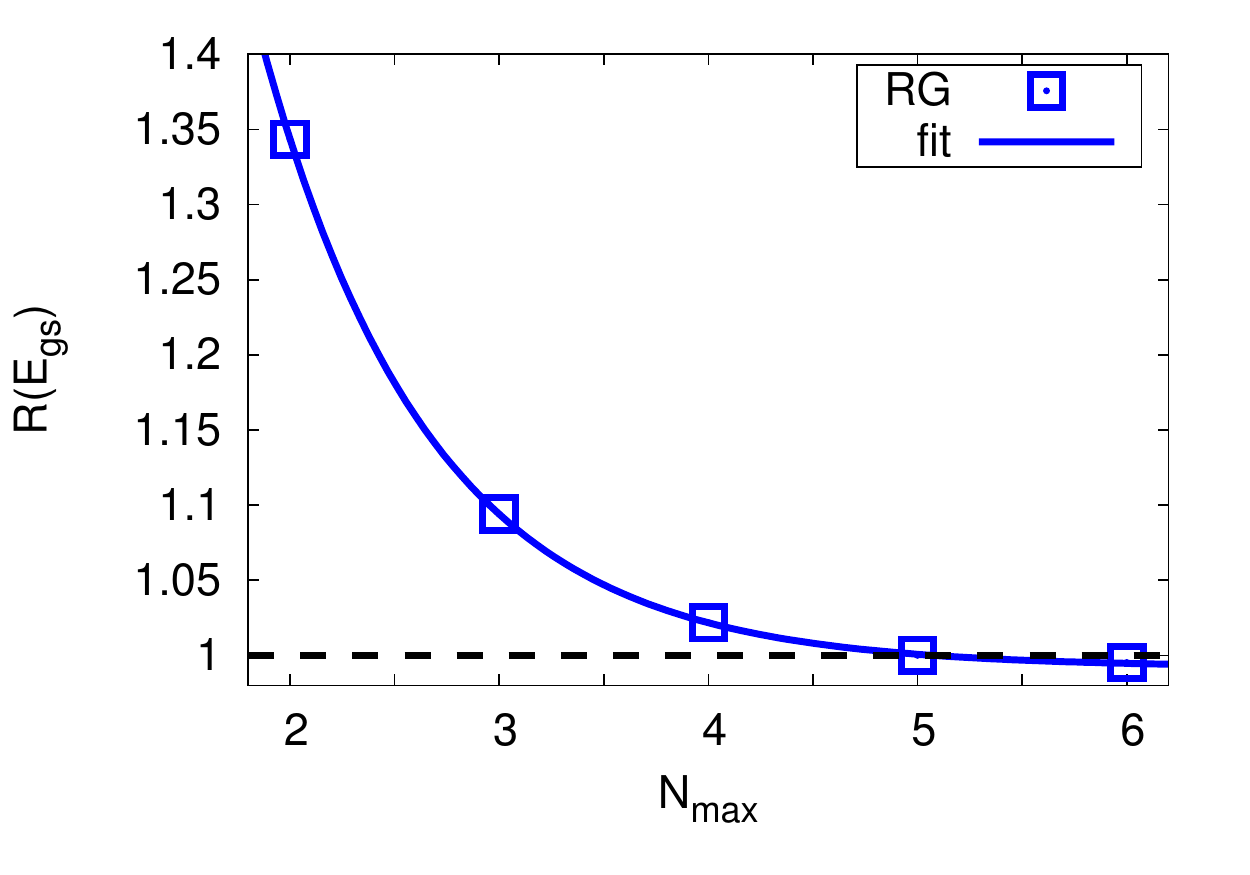}
\caption{$R(E_{\rm gs})=E_{\rm gs}^{\rm RG}/E_{\rm gs}^{\rm exact}$ as a function of~$N_{\rm max}$ for~\mbox{$\bar{u}_0=1$}. The solid (blue) line represents
a fit of the RG data to the (empirical) ansatz~\eqref{eq:empfitansatz}, see main text for details.}
\label{fig:egsd0conv}
\end{figure}

In Fig.~\ref{fig:egsd0_rg}, we present our RG results for~$E_{\rm gs}$ as a function of~$\bar{u}_0$
for various values for~$N_{\rm max}$. We find that our results for~$N_{\rm max}=4$ are already
in very good agreement with the exact results for~$\bar{u}_0 \lesssim 1$ and that we approach the exact
results from above for increasing~$N_{\rm max}$. For large values of the coupling~$\bar{u}_0$,
we observe that we need to go to higher truncation orders in order to reproduce the exact results.
Assuming that we do not know the exact solution for a given value of~$\bar{u}_0$, these findings imply
that we have to compute~$E_{\rm gs}$ as a function of~$N_{\rm max}$ and check numerically
the convergence of this function. This is illustrated  for~$\bar{u}_0=1$ in Fig.~\ref{fig:egsd0conv}. The 
solid (blue) line shows the result from a fit of 
the RG data for~$N_{\max}=2,3,4,5,6$ to the {(empirical)} three-parameter ansatz
\be
E_{\rm gs}^{\rm RG}(N_{\max}) = E_{\rm gs}^{\rm fit} + \alpha_1 {\rm e}^{-\alpha_2 N_{\rm max}}\,.
\label{eq:empfitansatz}
\ee
Here,~$E_{\rm gs}^{\rm fit}$, $\alpha_1$, and~$\alpha_2$ are
the three fit parameters. We obtain~$E_{\rm gs}^{\rm fit}\approx 0.0833$ which is about~$2\%$ smaller
than the exact result~$E_{\rm gs}\approx 0.0846$.

\subsection{Quantum Anharmonic Oscillator: One-dimensional Toy Model}\label{sec:tmaho}
Up to this point, we have only discussed a zero-dimensional model which essentially boils down to ordinary
calculus. Let us now discuss a simple one-dimensional field-theoretical toy model described by the following classical action:
\be
  && 
\!\!\!\!\!\!\!\!\!\!\!\! S = \frac{1}{2}\int _0^{\beta}d\tau\, \psi(\tau)\left[-\partial_{\tau}^2+\om^2\right]\psi(\tau) 
  \nn\\
  && 
   + \frac{1}{2} \int_0^{\beta}\! d\tau\! \int_0^{\beta}\! d\tau' \psi(\tau)\psi(\tau^{\prime})U(\tau\!-\!\tau')\psi(\tau^{\prime})\psi(\tau)\,,
   \label{eq:acttoymodeld1}
\ee 
where~$\psi(\tau)$ is a real-valued field and
\be
U(\tau-\tau^{\prime}) = \frac{1}{12}u_0\delta(\tau-\tau^{\prime})\,.
\ee
From a quantum mechanical point of view, this classical action describes nothing but the quantum 
anharmonic oscillator.\footnote{From a field-theoretical point of view, it corresponds to a so-called $\phi^4$-theory in one dimension.} 
Note that~$\psi(\tau)$ can be associated with a time-dependent 
coordinate. We have chosen the normalization factor of the interaction potential~$U$
such that it corresponds to the standard choice for this model, see, e.~g., Ref.~\cite{ZinnJustin}.

Much is known about this model. In fact, it is straightforward to diagonalize the Hamilton operator of this
model numerically without any approximation, see, e.~g., Ref.~\cite{Leonhardt}. 
Thus, we have again an exact solution at hand which allows us to benchmark our RG results. We would like to add that this 
model has already been employed to benchmark other RG approaches, such as the $1$PI RG 
approach~\cite{Horikoshi:1998sw,Gies:2006wv} and also 2PI approaches~\cite{Nagy:2010fv}. 
The large-coupling limit of this model has been studied using so-called
large-$N$ techniques, see, e.~g., Ref.~\cite{ZinnJustin}. In the latter approach, it is straightforward to show
that the ground-state energy scales as
\be
E_{\rm gs}\sim \omega \bar{u}_0^{\frac{1}{3}} \label{eq:egsln}
\ee
for~$\bar{u}_0 \gg 1$. Here,~$\bar{u}_0$ is the dimensionless coupling constant which is 
defined as\footnote{Note that the dimension of the coupling in units of~$\omega$ differs from the
one discussed in Eq.~\eqref{eq:dimcouplgen}. This can essentially be traced back to the fact that the time derivative
appears quadratically in the action~\eqref{eq:acttoymodeld1} of our present model rather than linearly.} 
\be
\bar{u}_0=\frac{u_0}{\omega^3}\,.
\ee

In the following we use our $2$PPI RG approach (DFT-RG approach) to study the ground-state properties
of this model. To this end, it is convenient to expand the ground-state~$\rho_{\rm gs,\lambda}(\tau)$ and the $n$-point
function~$\Gamma^{(n)}_{\lambda}[\rho_{\rm gs,\lambda}]$ in a Fourier series. For the ground-state, we choose
\be
\rho_{\rm gs,\lambda}(\tau)=\frac{1}{\beta}\sum_{n=-\infty}^{\infty} 
\rho_{\lambda}^{(n)} {\rm e}^{{\rm i}\omega_{n}\tau}\,,\label{eq:rfe}
\ee
where~$\omega_{n}=2\pi n/\beta$. Note that the fields~$\psi(\tau)$ are real-valued fields and therefore obey
periodic boundary conditions in the imaginary-time direction. 

Rather than working with the two-point function~$\Gamma^{(2)}_{\lambda}$,
we shall use the propagator~$G_{\lambda}(\tau,\tau^{\prime})$ from now on. The latter is defined as the inverse
of the two-point function:
\be
\!\!\!\int_0^{\beta} d\tau^{\prime\prime} G_{\lambda}(\tau,\tau^{\prime\prime})\left[ \Gamma^{(2)}_{\lambda}[\rho_{\rm gs,\lambda}](\tau^{\prime\prime},\tau^{\prime})\right]
=\delta (\tau\!-\!\tau^{\prime})\,.
\ee
Note that~$G_{\lambda}$ should not be confused with a single-particle propagator. 
However, both propagators are related in simple terms, see our detailed discussion in Appendix~\ref{sec:ic}.
The propagator can again be expanded in a Fourier series:
\be
\!\!\! G_{\lambda}(\tau,\tau^{\prime})\!=\!\frac{1}{\beta}\!\sum_{m=-\infty}^{\infty}\frac{1}{\beta}\!\sum_{n=-\infty}^{\infty}
G_{\lambda}^{(m,n)} {\rm e}^{{\rm i}\omega_{m}\tau} {\rm e}^{-{\rm i}\omega_{n}\tau^{\prime}}. \label{eq:gfe}
\ee
The matrix~$G_{\lambda}^{(m,n)}$ is diagonal, i.~e.
\be
G_{\lambda}^{(m,n)} = \beta G_{\lambda}^{(m)}\delta_{m,n}\,.
\ee
Higher $n$-point functions can be expanded accordingly:
\be
&&
\!\!\!\!\!\!\!\!\! \Gamma^{(n)}_{\lambda}[\rho_{\rm gs,\lambda}](\tau_1,\dots,\tau_n)
\nn\\
&& 
\!\!\!\!\! = \frac{1}{\beta^n}\!\sum_{m_1,\dots,m_n=-\infty}^{\infty}
\!\!\!\left(\Gamma^{(n)}_{\lambda}\right)^{(m_1,\dots,m_n)}
{\rm e}^{{\rm i}\sum_{l=1}^{n}\omega_{m_l}\tau_l}\,.
\ee
In order to derive the flow equations for the Fourier coefficients, we again employ our vertex 
expansion, see Eq.~\eqref{eq:vertexexp}. Plugging the latter into the general flow equation~\eqref{eq:DFTfloweq}, 
we eventually obtain the flow equations for the Fourier coefficients. The initial conditions for these
RG equations are determined by the non-interacting problem ($\bar{u}_0=0$), see Appendix~\ref{sec:ic}.
\begin{figure}[t]
\includegraphics[width=\linewidth]{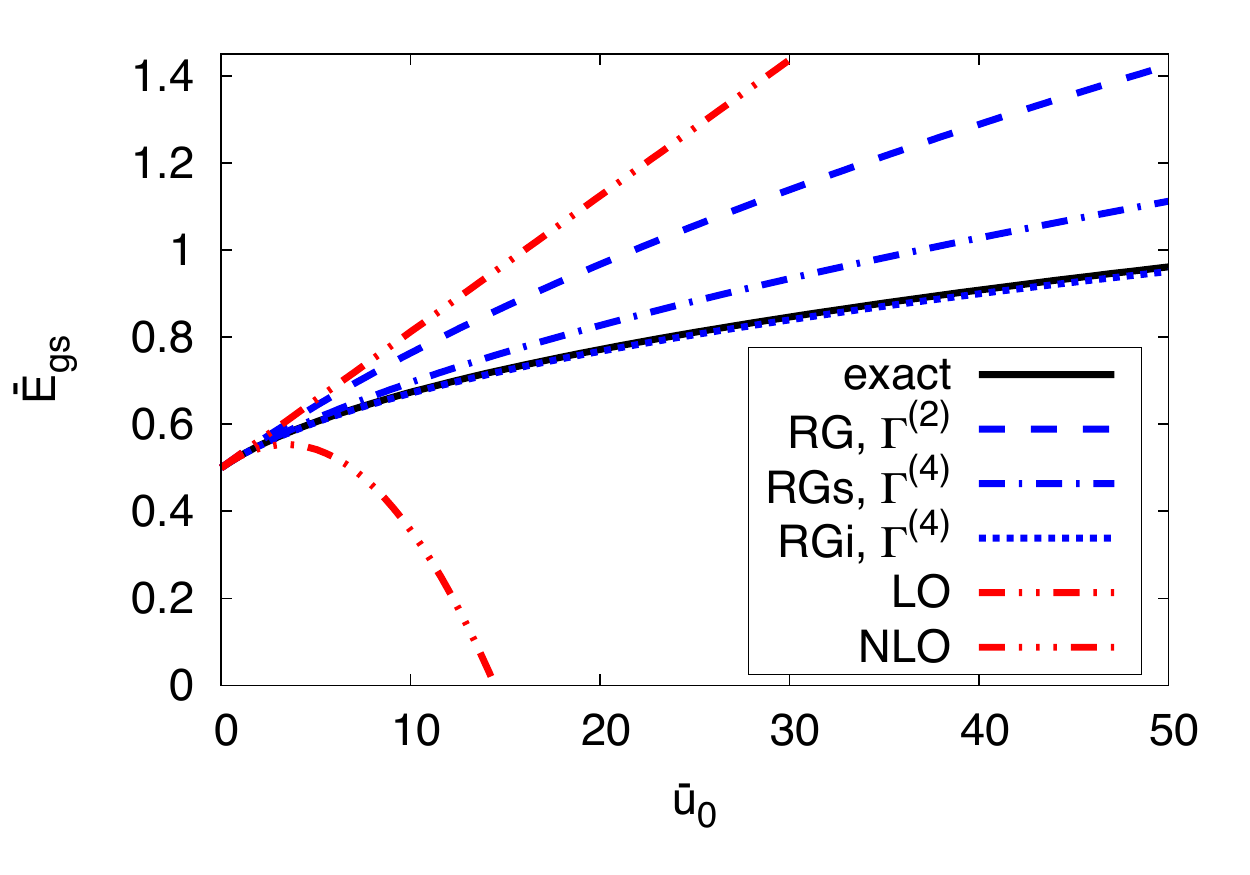}
\caption{
Comparison of the ground-state energy~$E_{\rm gs}$ as a function of~$\bar{u}_0$ as obtained from
various different approaches. The RG results are compared with the exact results and the perturbative 
results at leading order (LO) and next-to-leading order (NLO). The analytic solution~\eqref{eq:egsd1res}
for the leading-order RG approximation (RG, $\Gamma^{(2)}$) overestimates significantly the ground-state energy. 
For details on the RG approximations of higher order (RGs, $\Gamma^{(4)}$; RGi, $\Gamma^{(4)}$), we refer the reader to the
main text.
}
\label{fig:egsd1}
\end{figure}
\subsubsection{Leading-order approximation}
Let us begin with a discussion of the lowest non-trivial approximation which is given by dropping~$\Gamma^{(n)}_{\lambda}[\rho_{\rm gs,\lambda}]$
as well as their RG flows for~$n\geq 3$, i.~e. we consider the case $N_{\rm max}=2$. For the ground-state energy~$E_{\rm gs,\lambda}$, 
we then find
\be
  \partial_\lambda \bar{E}_{\rm gs,\lambda}
  =\frac{\bar{u}_0}{24} \left[  \left(\frac{1}{\bar{\beta}}\bar{\rho}^{(0)}_{\lambda}\right)^2 
  + \frac{1}{\bar{\beta}}\sum_{m=-\infty}^{\infty} \bar{G}_\lambda^{(m)}\right]\,. \label{eq:egsflowd1}
\ee
Here, we have introduced the following dimensionless quantities:
\be
\bar{E}_{\rm gs,\lambda}&=&\omega^{-1}E_{\rm gs,\lambda}\,,\\
\bar{\beta} &=& \beta\omega\,,\\
\bar{\rho}^{(m)}_{\lambda} &=& \omega^2 \rho^{(m)}_{\lambda}\,,\\
\bar{G}^{(m)}_{\lambda} &=& \omega^3 G^{(m)}_{\lambda}\,.
\ee
In terms of these quantities, 
the flow equations for the Fourier coefficients associated with the ground state~$\rho_{\rm gs}$ read
\be
  \partial_\lambda \bar{\rho}^{(m)}_{\lambda} 
  =-\frac{1}{12} \bar{u}_0\bar{\rho}^{(m)}_{\lambda} \bar{G}_\lambda^{(-m)}\,.\label{eq:rflowd1}
\ee
Finally, the RG flow of the propagator is determined by the following set of equations:
\be
 \partial_\lambda \bar{G}^{(m)}_\lambda=-\frac{1}{12}\bar{u}_0 \left(\bar{G}_\lambda^{(m)}\right)^2\,.\label{eq:gflowd1}
\ee
Clearly, Eqs.~\eqref{eq:egsflowd1},~\eqref{eq:rflowd1} and~\eqref{eq:gflowd1} represent an infinite set of flow equations.
For the present leading-order approximation, we can still solve this set analytically. We find
\be
\bar{E}_{\rm gs,\lambda}=-\frac{1}{2}+\frac{\bar{u}_0\lambda}{4(24+\bar{u}_0\lambda)}+\sqrt{1+\frac{\bar{u}_0}{24}\lambda}\,\,.
\label{eq:egsd1res}
\ee
The first two terms essentially represent the Hartree term whereas the last term originates from the third term
on the right-hand side of Eq.~\eqref{eq:DFTfloweq}.
As it should be, this expression reduces to~$\bar{E}_{\rm gs,\lambda=0}=\frac{1}{2}$ at the non-interacting
starting point of the flow ($\lambda=0$). For small~$\bar{u}_0$, we can expand the result for the ground-state energy
and obtain
\be
\bar{E}_{\rm gs,\lambda=1}=\frac{1}{2}+\frac{1}{32}\bar{u}_0 - \frac{1}{1536}\bar{u}_0^2 + {\mathcal O}(\bar{u}_0^3)\,,
\ee
which needs to be compared to the exact perturbative results~\cite{ZinnJustin}:
\be
\bar{E}_{\rm gs}^{\rm exact}= \frac{1}{2} +\frac{1}{32}\bar{u}_0 - \frac{7}{1536}\bar{u}_0^2 + {\mathcal O}(\bar{u}_0^3)\,.
\ee
Thus, we reproduce the perturbative result at leading order within the present approximation but not the coefficient of the 
second-order correction. Moreover, it follows from Eq.~\eqref{eq:egsd1res} that~$\bar{E}_{\rm gs}\equiv\bar{E}_{\rm gs,\lambda=1}$
scales as
\be
\bar{E}_{\rm gs} \sim \sqrt{\bar{u}_0}
\ee
for~$\bar{u}_0\gg 1$. Apparently, this does not agree with the result from the large-$N$ approximation, see Eq.~\eqref{eq:egsln}. The 
scaling behavior of the latter has also been confirmed for the present model by solving the Schr\"odinger equation numerically, see, e.~g., Ref.~\cite{Leonhardt}.
Note that the Hartree approximation renders~$E_{\rm gs}$ independent of~$\bar{u}_0$ in the large coupling limit.
\begin{figure}[t]
\includegraphics[width=\linewidth]{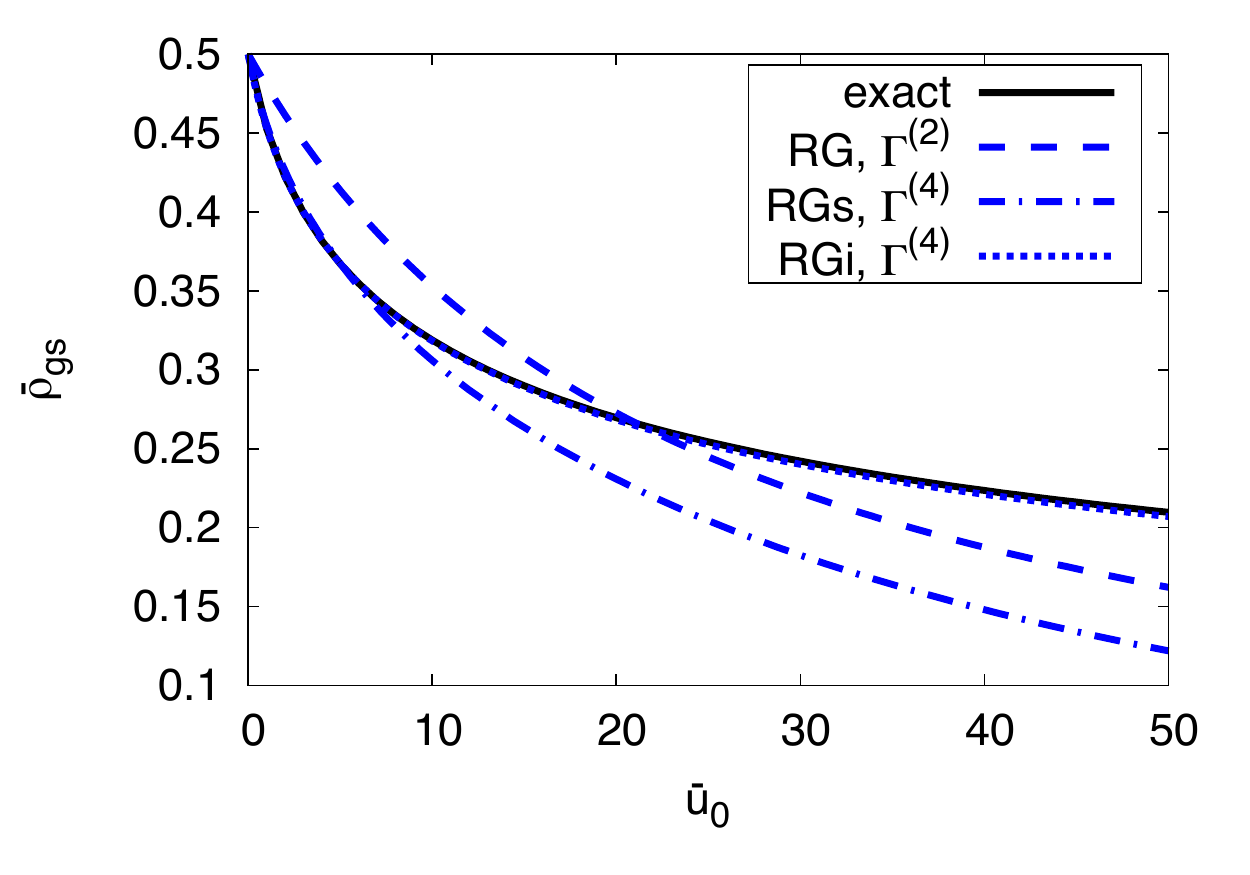}
\caption{
Comparison of the ground-state~$\rho_{\rm gs}$ as a function of~$\bar{u}_0$ as obtained from
various different approaches. The analytic solution~\eqref{eq:rgsd1res}
for the leading-order RG approximation (RG, $\Gamma^{(2)}$) is not in accordance with the exact results, neither in the small coupling-limit
nor for large values of the coupling,
see main text for details on the higher-order RG approximations (RGs, $\Gamma^{(4)}$; RGi, $\Gamma^{(4)}$).
}
\label{fig:rgsd1}
\end{figure}

In Fig.~\ref{fig:egsd1}, we show our results for the ground-state energy as a function of the dimensionless
coupling~$\bar{u}_0$. For comparison, we also show the exact results~\cite{Leonhardt} and the perturbative results at leading
and next-to-leading order. Our RG results  for the ground-state energy from the leading-order approximation is in agreement with
the leading-order perturbative result at small coupling but overestimates significantly
the exact results for large values of~$\bar{u}_0$. Thus, the present approximation is insufficient and 
we need to take into account $n$-point functions of higher order, see our discussion below.

For completeness, we also discuss the behavior of the ground-state as obtained from the present approximation.
First, we note that~$\bar{\rho}_{\lambda}^{(n)}=0$ for $\lambda=0$ and~$n\neq 0$, see also Appendix~\ref{sec:ic}. 
For~$n=0$, we have~$\bar{\rho}_{\lambda=0}^{(0)}\to\bar{\beta}/2$ for~$\bar{\beta}\to\infty$.
This implies that~$\partial_{\lambda}\rho_{\lambda}^{(n)}=0$ for~$n\neq 0$. Thus, only the zero mode~$\bar{\rho}_{\lambda}^{(0)}$
contributes to the flow of~$\rho_{\rm gs,\lambda}$: 
\be
\rho_{\rm gs,\lambda}(\tau) 
 \; \stackrel{\bar{\beta}\to\infty}{\longrightarrow}  \;
  \frac{1}{2\omega}\frac{1}{1+\lambda\frac{\bar{u}_0}{24}}\,.\label{eq:rgsd1res}
\ee
Note that~$\rho_{\rm gs,\lambda}(\tau)$ does not depend on the imaginary time~$\tau$.

Our results for~$\rho_{\rm gs,\lambda}(\tau)$ are shown in Fig.~\ref{fig:rgsd1}. Since the three-point
function~$\Gamma^{(3)}_{\lambda}$ contributes to the flow of the ground state~$\rho_{\rm gs}$ already
at leading order, we do not even recover the leading-order perturbative result with our present
truncation. For example, this becomes apparent from the comparison with the exact result in the small
coupling-limit. The same line of arguments holds for the propagator. In fact, 
not only the three-point function~$\Gamma^{(3)}_{\lambda}$ contributes to the flow of the propagator but also
the four-point function~$\Gamma^{(4)}_{\lambda}$.

\subsubsection{Higher-order approximations}

Let us now discuss the role of $n$-point functions of higher order. As discussed in Sect.~\ref{sec:RGDFT}, 
it can be shown that the RG flow of the $n$-point function depends on the flow of the~$n+1$ and $n+2$-point function.
Since the ground-state is associated with the one-point function,\footnote{The flow equation for the ground state follows
from the stationary condition~$(\delta\Gamma/\delta\rho)|_{\rm gs}=0$.} this means that the two- and the three-point function
govern the RG flow of~$\rho_{\rm gs}$. In fact, the most general form of the flow equation for the ground state~$\rho_{\rm gs}$ of our present 
model with a constant background potential is given {by
\be
&&  \!\!\!\!\!\!\!\!\!\!
\partial_\lambda \bar{\rho}^{(l)}_{\lambda} 
= - \frac{\bar{u}_0}{24} \Bigg[ 2\bar{\rho}_{\lambda}^{(l)} \bar{G}_\lambda^{(-l)}
 \nn\\
 && \quad\qquad 
 -\frac{\delta_{l,0}}{\bar{\beta}}\sum_{k=-\infty}^{\infty}\!\!
 \bar{G}_\lambda^{(l)}\left(\bar{G}_\lambda^{(k)}\right)^2\left(\bar{\Gamma}_\lambda^{(3)}\right)^{(k,-k,l)} \Bigg].\label{eq:flowrhogen}
 \\ \nn
\ee
Since}~$\bar{\rho}_{\lambda}^{(n)}=0$ for $\lambda=0$ and~$n\neq 0$, only the flow of the zero mode~$\bar{\rho}_{\lambda}^{(0)}$ 
is {non-}vanishing and therefore only~$\bar{\rho}_{\lambda}^{(0)}$ assumes a finite value. From this observation, it follows
immediately that the ground state~$\rho_{\rm gs}$ does not depend on the imaginary time. This remains true even for
$d\!+\!1$-dimensional field theories, provided that we consider an interaction potential of the form~$U\sim \delta(\tau-\tau^{\prime})$.
In any case, from the general flow equation for the ground state it is apparent that the three-point function
contributes to the ground-state at leading order in a perturbative expansion.

Let us now turn to the most general flow equation for the propagator (inverse two-point function). This flow equation depends
on the RG flows of the three- and four-point function and can be written as follows
\begin{widetext}
\be
&& \partial_\lambda \bar{G}^{(m)}_\lambda= -\frac{\bar{u}_0}{12} \Bigg[
  \left(\bar{G}_\lambda^{(m)}\right)^2 
  \!\!-\!  \frac{1}{\bar{\beta}^2}\sum_{l=-\infty}^{\infty}\! \bar{\rho}_{\lambda}^{(l)} \bar{G}_\lambda^{(-l)}\!\left(\bar{G}_\lambda^{(m)}\right)^2\!\left(\bar\Gamma_\lambda^{(3)}\right)^{(-l,m,-m)}
\! \!-\!\frac{1}{6} \frac{1}{\bar{\beta}^2}\sum_{k=-\infty}^{\infty}\!\!\left(\bar{G}_\lambda^{(m)}\right)^2\!\left(\bar{G}_\lambda^{(k)}\right)^2\!
 \left(\bar\Gamma_\lambda^{(4)}\right)^{(k,m,-k,-m)} 
 \nn\\
&&  
\qquad\qquad\qquad\qquad\qquad +\frac{1}{2} \frac{1}{\bar{\beta}^3}\sum_{l=-\infty}^{\infty}
\sum_{k=-\infty}^{\infty}\bar{G}_\lambda^{(-l)}\left(\bar{G}_\lambda^{(m)}\right)^2\left(\bar{G}_\lambda^{(k)}\right)^2\left(\bar\Gamma_\lambda^{(3)}\right)^{(k,-k,l)}\left(\bar\Gamma_\lambda^{(3)}\right)^{(-l,m,-m)}  
\nn\\
 && 
 \qquad\qquad\qquad\qquad\qquad\qquad + \frac{1}{\bar{\beta}^3}\sum_{l=-\infty}^{\infty}\sum_{k=-\infty}^{\infty}
 \left(\bar{G}_\lambda^{(m)}\right)^2\left(\bar{G}_\lambda^{(k)}\right)^2\bar{G}_\lambda^{(m+k)}\left(\bar\Gamma_\lambda^{(3)}\right)^{(k,m,l)}\left(\bar\Gamma_\lambda^{(3)}\right)^{(-l,-m,-k)} 
 \nn\\
 && 
 \qquad\qquad\qquad\qquad\qquad\qquad\qquad -\frac{1}{3}  \frac{1}{\bar{\beta}^2}\sum_{k=-\infty}^{\infty}
 \left(\bar{G}_\lambda^{(m)}\right)^2\left(\bar{G}_\lambda^{(k)}\right)^2\left(\bar\Gamma_\lambda^{(4)}\right)^{(k,-k,m,-m)} \Bigg]\,.
\label{eq:flowpropgen}
\ee
\end{widetext}
Since the right-hand side is proportional to the coupling~$\bar{u}_0$, we 
conclude that we need to include the three- and four-point function in our study in order to recover the correct 
perturbative result at leading order. The inclusion of these two correlation functions in the flow of the propagator
then also guarantees that the results for the ground-state energy agree with perturbation theory up to second order
in the small-coupling limit, see our general discussion in Sect.~\ref{sec:pt}. 

For simplicity, we do not include the full flow of the three- and four-point function in our numerical studies
in the following. We rather use
a `static' approximation for these correlation functions, i.~e. we set their flow equations to zero, 
$\partial_{\lambda}\Gamma^{(3)}_{\lambda}\!=\!\partial_{\lambda}\Gamma^{(4)}_{\lambda}\!=\!0$, but we do not set the functions 
themselves to zero.
Recall that~$\Gamma^{(3)}_{\lambda}$ and~$\Gamma^{(4)}_{\lambda}$ are not identical to zero at the 
initial point of the RG~flow ($\lambda=0$), see also Appendix~\ref{sec:ic}. 
Moreover, we set the $n$-point correlation functions with $n\!>\!4$ to zero as well as their flows.

Using this approximation, we recover the correct leading-order behavior of the propagator and the ground-state~$\rho_{\rm gs}$ 
in the small-coupling limit. Moreover, the ground-state energy is correct up to second order in this limit, as
we have checked analytically. 

In order to go beyond the perturbative small-coupling limit, we have to solve the set of flow 
equations~\eqref{eq:egsflowd1},~\eqref{eq:flowrhogen}, and~\eqref{eq:flowpropgen} numerically. To this end,
we have to truncate the sum over the Fourier coefficients, i.~e. we only take into account the coefficients up to a certain
value~$N_{\rm max}^{\rm Fourier}$. Here and in our subsequent study including an RG improvement, we choose~$N_{\rm max}^{\rm Fourier}=100$
{and~$(\bar{\beta})^{-1}=T/\omega=0.1$ for} the dimensionless temperature. We have also applied this choice to the lowest-order RG approximation and
compared the results to the corresponding analytic solution. For~$\bar{u}_0\lesssim 50$, 
we have found that our numerical results for~$E_{\rm gs}$ and~$\rho_{\rm gs}$ deviate from the analytic ones by about~$1\%$
or less.\footnote{For increasing~$\bar{u}_0$, we have to increase $N_{\rm max}^{\rm Fourier}$ in order to reproduce
the analytic results.}

In Fig.~\ref{fig:egsd1}, we present our numerical results for the ground-state energy obtained with this approximation (labelled as `RGs, $\Gamma^{(4)}$'). The
ground-state is shown in Fig.~\ref{fig:rgsd1}. We observe that the ground-state energy is now in reasonable agreement
with the exact result. To be more specific, we find that the ground-state energy is less than $5\%$ larger than the exact value 
at~$\bar{u}_0=10$. For~$\bar{u}_0=20$, however, we still have an error of about~$10\%$. In any case, the inclusion 
of~$\Gamma^{(3)}_{\lambda}$ and~$\Gamma^{(4)}_{\lambda}$ in our static approximation already yields a drastic improvement 
of our results obtained with the leading-order approximation.
With respect to the ground-state, we now find good agreement
between our results and the exact results for~$\bar{u}_0\lesssim 2$.  
For larger values of the coupling, we observe that our results deviate
significantly from the exact results. In fact, also the asymptotic functional form of~$\rho_{\rm gs}$ turns out to be incorrect.

\subsubsection{RG improvement}

Let us now discuss how our RG flows can be improved. To this end, we consider the approximation which we have just discussed.
The results for~$E_{\rm gs}$ from this approximation are already in reasonable agreement with the exact result for~$\bar{u}_0 \lesssim 20$. For larger values
of the coupling, this approximation suffers from the fact that we have only included~$\Gamma^{(3)}_{\lambda}$ and~$\Gamma^{(4)}_{\lambda}$
`statically', i.~e. we have dropped the flow of these correlation functions. Now we would like to improve our set of flow equations without
explicitly taking into account the RG flow equations for~$\Gamma^{(3)}_{\lambda}$ and~$\Gamma^{(4)}_{\lambda}$. 

First, we note that 
the oscillator frequency~$\omega$ effectively changes when the interaction is turned on.
To be more specific,~$\omega$ effectively increases with
increasing~$\bar{u}_0$ due to quantum corrections. This becomes apparent when we look at the $\bar{u}_0$-dependence of 
the ground state~$\rho_{\rm gs}$. 
At~$\lambda=0$ (non-interacting limit), we have
\be
\rho_{\rm gs,\lambda=0}=\frac{1}{2\omega}
\ee
for~$\beta\to\infty$. Now recall that for increasing~$\lambda$ and/or~$\bar{u}_0$, the 
value for the ground state $\rho_{{\rm gs},\lambda}$ decreases, see also Eq.~\eqref{eq:rgsd1res}.
Thus, a change of the ground state can therefore be viewed as an effective change of the oscillator frequency.
In this spirit, we define an effective $\lambda$-dependent oscillator frequency~$\omega^{\rm eff.}_{\lambda}$:
\be
\omega_{\lambda}^{\rm eff.} := \frac{1}{2}\left(\rho_{\rm gs,\lambda}\right)^{-1}\,.
\ee
In terms of the only non-vanishing Fourier coefficient of the ground-state, this can be written as follows:
\be
\bar{\omega}_{\lambda}^{\rm eff.} =  \frac{\omega_{\lambda}^{\rm eff.}}{\omega}= \frac{\bar{\beta}}{2}\left(\bar{\rho}_{\rm gs,\lambda}^{(0)}\right)^{-1}\,.
\ee
With this effective oscillatory frequency at hand, we can improve our set of RG equations for the ground-state and the 
propagator. 
To this end we exploit the fact that~$\Gamma^{(3)}_{\lambda=0}$ and~$\Gamma^{(4)}_{\lambda=0}$ 
depend on the oscillatory frequency~$\omega$. We now replace~$\omega$ with the 
effective $\lambda$-dependent (flowing) oscillatory frequency~$\omega_{\lambda}^{\rm eff.}$. 
This can be viewed as 
an inclusion of the three- and four-point function associated with a harmonic oscillator with frequency~$\omega_{\lambda}^{\rm eff.}$. 
Since~$\omega_{\lambda}^{\rm eff.}$ is a non-trivial function of~$\bar{u}_0$, this improvement allows us to
include corrections of higher order in a simple manner.
 
We would like to add that the change of~$\Gamma^{(3)}_{\lambda}$ and~$\Gamma^{(4)}_{\lambda}$ under a variation of~$\lambda$
is in principle governed by flow equations which we have not taken into account in the present study. In this respect, our 
improvement can also be considered as a way to estimate the impact of neglected $n$-point functions of higher order. 
Recall that the flow of a given $n$-point function depends on the $n+1$- and $n+2$-point function. This implies that
we have in general to deal with an infinite tower of flow equations. Even 
if we had included the full flow equations for the three- and four-point function, we would still require information about 
the five- and six-point function. Since this tower of equations needs to be truncated at some order, it is always desirable to
estimate to some extent the effect of the neglected $n$-point functions.
Our proposed RG improvement represents
one possibility for such an estimate.

Let us now turn to the (numerical) results from our improved RG flows. From the construction of these flows, it is clear that
we still recover the correct perturbative result for the ground-state energy 
up to second order. To see this, we note 
that~$\omega_{\lambda}^{\rm eff.}$ can be expanded in a power series of~$\bar{u}_0$. Our improved
three- and four-point functions therefore also have a well-defined perturbative expansion:
\be
&& \Gamma^{(n)}_{\lambda=0}\Big|_{\omega = \omega_{\lambda}^{\rm eff.}} - \Gamma^{(n)}_{\lambda=0}\Big|_{\omega} \nn\\
&& \qquad\; = \left( \frac{\partial\omega_{\lambda}^{\rm eff.}}{\partial \bar{u}_0} \right)
 \left( \frac{\partial \Gamma^{(n)}_{\lambda=0}}{\partial \omega} \right)_{\omega = \omega_{\lambda}^{\rm eff.}} 
 \bar{u}_0 + {\mathcal O}(\bar{u}_0^2)\,.
\ee
From this expression, it is apparent that the improvement only affects the third order of the perturbative expansion of the RG flow equation
for the ground-state energy.
For the ground state~$\rho_{\rm gs}$, the second order already receives contributions from our RG improvement. The latter statement also holds
for the propagator.
In Fig.~\ref{fig:egsd1} we show our results for the ground-state energy obtained with this RG improvement (labelled as `RGi, $\Gamma^{(4)}$'). 
{The results for the ground-state are given in Fig.~\ref{fig:rgsd1}. We observe that the improvement brings our RG results for~$E_{\rm gs}$ and~$\rho_{\rm gs}$
very close to the corresponding exact values.
In fact, we find that both the ground-state energy as well as the ground-state~$\rho_{\rm gs}$ are almost indistinguishable from
the exact results on the scale of the plot for the depicted range of values for the coupling.
Finally,} we would like to add that our RG results for the ground-state energy
approach the exact results from above when we include $n$-point functions of higher order.
Comparing our results for~$\rho_{\rm gs}$ with the exact values, 
it even appears that our results are now consistent with the asymptotic functional form of~$\rho_{\rm gs}$ for large values of the coupling.

\section{Conclusions and Outlook}\label{sec:conc}

We have discussed an RG approach to DFT and analyzed its properties on very general grounds,
including its relation to perturbation theory and the computation of excited states. A special feature
of this approach is that it does not rely on a (global) parameterization of the density functional. The RG flow
rather starts at a well-defined analytically accessible starting point. By gradually increasing the {\it microscopic} interactions,
the underlying RG flow equation then allows us to follow the ground-state of the theory. 

The presented RG approach
eventually aims at a study of ground-state properties of 
non-relativistic self-bound fermionic systems, such as (heavy) nuclei, from microscopic interactions.
In the present work, however, we have restricted ourselves to studies of two simple toy models which 
nevertheless helped us to already test and benchmark our approach in a simple but meaningful manner
for future studies. From a theoretical point of view, these two models correspond to a zero 
and one-dimensional field theory. In the present paper, the latter was nothing but the quantum anharmonic oscillator.

Our toy model studies allowed us to illustrate how our RG approach works in general and how ground-state
properties of physical systems can be computed with this novel tool. An important feature of our approach
is that the so-called $2$PPI effective action (corresponding to the energy density functional) can be computed
systematically from the underlying microscopic interactions by means of a vertex expansion. With respect
to nuclear physics, this implies that our DFT-RG approach opens up a new direction to compute 
the energy density functional from, e.~g., chiral EFT interactions. Apart from the explicit (quantitative) computation
of energy functionals and ground-state properties, the presented RG approach can be viewed as a tool 
to gain deeper insights into the general structure of density functionals. For example, we have demonstrated
how a connection between the energy density functional and perturbation theory can be established
in a simple and systematic fashion.

For our toy models, we have found that the RG results for, e.~g., the ground-state energy 
are in reasonable agreement with the exact results 
over a wide range of values for the (microscopic) coupling constant, even if we only
take into account correlation functions up to the four-point function. By comparing our results with the exact results, 
we have also demonstrated that the quality of the RG results can be systematically improved by taking into account $n$-point
functions of higher order. 

{The present work should be considered as a starting point for various studies. For example, we have already mentioned the relation of the present
DFT-RG approach to the 1PI RG approach~\cite{Wetterich:1992yh}. It would also be
interesting to better understand how it relates to other RG approaches widely used to study
the nuclear many-body problem~\cite{Bogner:2006pc,Jurgenson:2008jp,Tsukiyama:2010rj,Hergert:2012nb}.
Apart from these more field-theoretically motivated formal studies, 
another natural next step} is now to study ground-state properties of so-called
Alexandrou-Negele nuclei~\cite{Alexandrou:1988jg}. These are self-bound systems in one space and one time dimension,
consisting of  $N$~spinless fermions interacting
via a specific choice for a long-range attractive and short-range repulsive potential~$U$. In this case, 
the initial condition at~$\lambda=0$ corresponds to a simple harmonic oscillator potential in which the $N$ lowest 
lying levels are filled. By solving the RG flow equation~\eqref{eq:DFTfloweq}, we then gradually remove the 
background potential~$V$ and turn on the two-body interaction potential $U$. The RG results
for these systems can be benchmarked against those from {\it ab-initio} MC calculations~\cite{Alexandrou:1988jg} and
studies with the so-called similarity RG approach~\cite{Jurgenson:2008jp}. The present work will
help us to set up these studies of one-dimensional nuclei~\cite{BSP}. The latter will then help us to further develop
and establish our novel DFT-RG approach and may pave the way for the computation of ground-state properties
of realistic nuclei from microscopic interactions in a novel alternative way.

\vspace{-0.3cm}
\acknowledgments
The authors would like to thank J.~Berges, T. Duguet, 
R.~J.~Furnstahl, H.~Gies, F.~Karbstein, T.~Papenbrock, 
J.~M.~Pawlowski, J.~Polonyi, and A.~Schwenk for helpful discussions.
Moreover, the authors are very grateful to R.~J. Furnstahl, J.~Polonyi, and A.~Schwenk for useful comments on the manuscript.
In addition, we would like to thank J.~F.~Rentrop, S.~G.~Jakobs, and V.~Meden for numerous discussions hinting us to 
a typo in the {\it numerical} implementation of the approximation ``RGi, $\Gamma^{(4)}$" (see Figs.~\ref{fig:egsd1} and~\ref{fig:rgsd1})
which affects the results in this very approximation in the case of strong couplings.
JB acknowledges support by the Helmholtz International Center for FAIR within the LOEWE program of the State of Hesse.
Moreover, the authors acknowledge support by the Deutsche Forschungsgemeinschaft (DFG) through contract SFB~634.


%
\appendix
\section{Initial Conditions}\label{sec:ic}
In this appendix we discuss the computation of the initial conditions for our one-dimensional toy model.
In the spirit of our vertex expansion, we expand the effective action about the ground state~$\rho_{\rm gs}$, see Eq.~\eqref{eq:vertexexp}.
Along these lines we also expand the source~$J$ about~$\rho_{\rm gs}$:
\be
&& \!\!\! J [\rho] =  
\int _{\tau^{\prime}}\, j_1(\tau,\tau^{\prime})[\rho(\tau^{\prime})\!-\! \rho_{\rm gs}(\tau^{\prime})]
\nn\\
&&
\qquad\qquad +\frac{1}{2} \int _{\tau^{\prime}}  \int_{\tau^{\prime\prime}}
\, j_2(\tau,\tau^{\prime},\tau^{\prime\prime}) [\rho(\tau^{\prime})\!-\! \rho_{\rm gs} (\tau^{\prime})]\times\nn\\
&&\qquad\qquad\qquad\quad\qquad \times[\rho(\tau^{\prime\prime})\!-\!\rho_{\rm gs}(\tau^{\prime\prime})]
+\dots\,. \label{eq:jexp}
\ee
Recall that~$J$ is a function of~$\tau$ but also a functional of~$\rho$. At the physical ground state~$\rho_{\rm gs}$, we 
have~$J[\rho=\rho_{\rm gs}]=0$. Using Eqs.~\eqref{eq:statcond} and~\eqref{eq:vertexexp}, we then find 
\be
j_1 = {\Gamma}^{(2)}_0\,, \qquad j_2 = {\Gamma}^{(3)}_0\,, \qquad \dots\,.
\ee
Here and in the following we use the index `0' (corresponding to $\lambda=0$) to indicate that we are only discussing the non-interacting limit
which determines the initial conditions for our RG flow equations.

For a non-interacting theory, the functional~$W_0$ is given by
\be
\!\!\!\!\!\!\! W_{0}[J]&=& -\frac{1}{2}\Tr \ln \left (\Delta_{0}^{-1} \!-\! 2\tilde{J}\right)\,\nn\\
& =& -\frac{1}{2} \Tr \ln\Delta_{0}^{-1} \!+\! \frac{1}{2}\sum_{n=1}^{\infty}\frac{1}{n}\Tr\left[ \left(2\Delta_{0}\cdot \tilde{J}\right)^{n}\right],
\label{eq:Wexp}
\ee
where $\Delta_{0}^{-1}$ denotes the (conventional) propagator and~$\tilde{J}$ is related to the source~$J$:
\be
\tilde{J}(\tau,\tau^{\prime})=J(\tau)\delta(\tau - \tau^{\prime})\,.
\ee
From a comparison of Eq.~\eqref{eq:Wexp} with the expansion~\eqref{eq:vertexexp} of the effective action, 
we immediately obtain~${\Gamma}_0$:
\be
{\Gamma}_{0}[\rho_{\rm gs}]= \frac{1}{2}\Tr \ln \Delta_{0}^{-1}\,.\label{eq:g0ni}
\ee
For the propagator (inverse density-density correlator), we find
\be
G_0(\tau,\tau^{\prime}) = 2 \Delta_{0} (\tau,\tau^{\prime})\Delta_{0} (\tau^{\prime},\tau) 
\,,\label{eq:G0}
\ee
where
\be
\Delta_{0}(\tau,\tau^{\prime})=\frac{1}{\beta}\sum_{n=-\infty}^{\infty}\frac{1}{\omega_{n}^{2} + \omega^{2}}\,{\rm e}^{{\rm i}\omega_{n}(\tau - \tau^{\prime})}\,.
\label{eq:spprop}
\ee
Note that~$\Delta_{0}(\tau,\tau^{\prime})=\Delta_{0}(\tau^{\prime},\tau)$. Inserting Eq.~\eqref{eq:spprop} into Eq.~\eqref{eq:g0ni} we 
obtain~$\Gamma_0[\rho_{\rm gs}]$ for the non-interacting system:
\be
{\Gamma}_{0}[\rho_{\rm gs}]=\frac{\beta\omega}{2}+\ln\left(1-e^{-\beta\omega}\right)\,.
\ee
From this expression, we can then extract the ground-state energy:
\be
E_{\rm gs} = \lim_{\beta\to\infty}\frac{1}{\beta}{\Gamma}_{0}[\rho_{\rm gs}]=\frac{\omega}{2}\,.
\ee
Next, we compute the Fourier expansion coefficients associated with the propagator~$G$ defined in Eq.~\eqref{eq:gfe}. Using
Eq.~\eqref{eq:G0}, we find 
\be
{G}_{0}^{(n)}=\frac{2}{\omega\left(\omega_n^2+4\omega^2\right)} \coth\left(\frac{\beta\omega}{2}\right) \,.
\ee
These coefficients serve as initial conditions for the RG flow of the propagator in our study of the one-dimensional toy model.

Let us now turn to the ground state~$\rho_{\rm gs}$ of the non-interacting theory. To this end,
we consider the relation between~$\rho$ and the functional~$W_0$:
\be
\rho(\tau)=\frac{\delta W_0}{\delta J(\tau)} &=& \Delta_0(\tau,\tau)\nn\\
&& \quad + 2 \int_0^{\beta} d\tau^{\prime} \Delta_0(\tau^{\prime},\tau)
\Delta_0(\tau,\tau^{\prime})J(\tau^{\prime})\nn\\
&& \quad\quad +\dots\,.
\ee
From this expression, we deduce that 
\be
\rho_{\rm gs}(\tau) = \Delta_{0}(\tau,\tau) = \frac{1}{2\omega}{\coth\left(\frac{\beta\omega}{2} \right)}\,.
\ee
It then follows that the Fourier expansion coefficients defined in Eq.~\eqref{eq:rfe} are given by 
\be
\rho_{\rm gs}^{(n)}= \frac{\beta}{2\omega}\coth\left(\frac{\beta\omega}{2} \right)\delta_{n,0}\,.
\ee
Moreover, we find
\be
&&\!\!\!\!\!\!\!\!\left(\rho(\tau) - \rho_{\rm gs}(\tau)\right) \nn\\
&& \quad = 2 \int _0^{\beta}d\tau^{\prime} \Delta_0(\tau^{\prime},\tau)
\Delta_0(\tau,\tau^{\prime})J(\tau^{\prime})+\dots\,.\label{eq:nnbarexp}
\ee
Plugging the expansion~\eqref{eq:jexp} into this expression, we obtain
\begin{widetext}
\be
&&\left(\rho(\tau) - \rho_{\rm gs}(\tau)\right) 
=
  \int _0^{\beta}d\tau^{\prime}
  G_{0} (\tau,\tau^{\prime})
 \bigg[ \int_0^{\beta} d\tau^{\prime\prime}\, j_{1}(\tau^{\prime},\tau^{\prime\prime})[\rho(\tau^{\prime\prime})-\rho_{\rm gs}(\tau^{\prime\prime})]
 \nn\\
&&  \qquad\qquad\qquad\qquad
+\frac{1}{2} \int _0^{\beta}d\tau^{\prime\prime}  \int _0^{\beta}d\tau^{\prime\prime\prime} 
\, j_{2}(\tau^{\prime},\tau^{\prime\prime},\tau^{\prime\prime\prime}) [\rho(\tau^{\prime\prime})-\rho_{\rm gs}(\tau^{\prime\prime})][\rho(\tau^{\prime\prime\prime})-\rho_{\rm gs}(\tau^{\prime\prime\prime})]
+\dots \bigg] 
 + \dots\,.
\ee
\end{widetext}
Here, we have used Eq.~\eqref{eq:G0} and~$G_0(\tau,\tau^{\prime})=G_0(\tau^{\prime},\tau)$.
In order to determine the expansion coefficients~$j_{n}$, we compare the left- and right-hand side order by order in
our functional Taylor expansion in powers of~$(\rho-\rho_{\rm gs})$. For the coefficient~$j_{1}$, we obtain
\be
\int d\tau^{\prime} G_{0} (\tau,\tau^{\prime}) j_{1}(\tau^{\prime},\tau^{\prime\prime}) = \delta (\tau-\tau^{\prime\prime})\,.
\ee
Thus, we have
\be
j_{1}(\tau,\tau^{\prime}) = G_{0}^{-1}(\tau,\tau^{\prime})= {\Gamma}^{(2)}_{0} (\tau,\tau^{\prime})\,.\label{eq:j1res}
\ee
Next, we would like to find a simple relation for~${\Gamma}^{(3)}$. This can be done along the lines
of our derivation~\eqref{eq:j1res}. However, we now need to expand
the right-hand side of Eq.~\eqref{eq:nnbarexp} up to second order in the source~$J$. From a comparison of the
expansion coefficients on the left- and right-hand side, we then find
\begin{widetext}
\be
&&\!\!\!\!\!\!\!\!\!\!\!\! {\Gamma}^{(3)}_{0}(\tau,\tau^{\prime},\tau^{\prime\prime})
= j_{2}(\tau,\tau^{\prime},\tau^{\prime\prime})
\nn\\
&& \!\!\!\!\!\!
= - 8\! \int _0^{\beta} \! d\tau^{\prime\prime\prime} \! \int _0^{\beta} \! d\tau^{\prime\prime\prime\prime} \! \int _0^{\beta} \! d\tau^{\prime\prime\prime\prime\prime}
G_{0}^{-1}(\tau,\tau^{\prime\prime\prime})
\Delta_{0} (\tau^{\prime\prime\prime},\tau^{\prime\prime\prime\prime}) 
G_{0}^{-1}(\tau^{\prime\prime\prime\prime},\tau^{\prime})\Delta_{0} (\tau^{\prime\prime\prime\prime},\tau^{\prime\prime\prime\prime\prime})
G_{0}^{-1}(\tau^{\prime\prime\prime\prime\prime},\tau^{\prime\prime})\Delta_{0} (\tau^{\prime\prime\prime\prime\prime},\tau^{\prime\prime\prime})\,,
\ee
\end{widetext}
where we have used Eq.~\eqref{eq:j1res}. The Fourier expansion coefficients of the three-point functions 
can be computed straightforwardly and read
\begin{widetext}
\be
\left({\Gamma}^{(3)}_{0}\right)^{(k,l,m)}&=&-\frac{\beta \omega^2}{\left(\coth\left(\frac{1}{2}{\beta\omega}\right)\right)^{2}} 
\left(\omega_k^2+\omega_k\omega_l 
+\omega_l^2+12\omega^2\right)\delta_{k+l+m,0}\,.
\ee
\end{widetext}

In our discussion of the one-dimensional toy model, we also
used the initial condition of the four-point function.
The computation of this correlation function (as well as of correlation functions of higher order) 
for the non-interacting theory can be done
following the procedure detailed above. For the coefficients of the Fourier expansion of the four-point
function we then obtain
\begin{widetext}
\be
\left({\Gamma}^{(4)}_0\right)^{(k,l,m,n)}&=& \frac{3\beta \omega^3}{\left(\coth\left(\frac{1}{2}{\beta\omega}\right)\right)^{3}}
\Big[\left(\omega_{k+l}^2\!+\!\omega_{k+l}\omega_m 
+\omega_m^2\!+\!12\omega^2\right)\left(\omega_k^2\!+\!\omega_k\omega_l \!+\! \omega_l^2\!+\! 12\omega^2\right)\left(\omega_{k+l}^2+4\omega^2\right)^{-1} 
 \nn\\
&& 
\qquad +\left(\omega_{l}^2+\omega_{l}\omega_m+\omega_m^2
+ 12\omega^2\right)\!\!\left(\omega_k^2\!+\!\omega_k\omega_{l+m}+\omega_{l+m}^2\!+\!12\omega^2\right)
\left(\omega_{l+m}^2+4\omega^2\right)^{-1}  \nn\\
&& 
\qquad\qquad -f(\omega_k,\omega_l,\omega_m,\omega)
\left(\left(\omega_{k+l}^2\!+\! 4\omega^2\right)\!\left(\omega_{l+m}^2\!+\! 4\omega^2\right)\right)^{-1}\Big]\delta_{k+l+m+n,0}\,,
\ee
 \end{widetext}
where
\begin{widetext}
 \be
 f(\omega_k,\omega_l,\omega_m,\omega)&=&\,640 \omega^6 
 + 48 \omega^4 \left[3 \omega_k^2 + 4 \omega_l^2 + 4 \omega_l \omega_m + 3 \omega_m^2 + 2 \omega_k (2 \omega_l + \omega_m)\right] \nn\\
&& + 4 \omega^2 \left[3 \omega_k^4 + 2 \omega_l^4 + 4 \omega_l^3 \omega_m + 10 \omega_l^2 \omega_m^2 + 8 \omega_l \omega_m^3
 + 3 \omega_m^4 + 4 \omega_k^3 (2 \omega_l + \omega_m) \right.\nn \\
 && \left. + 2 \omega_k (2 \omega_l + \omega_m) (\omega_l^2 + \omega_l \omega_m + 2 \omega_m^2) 
  + 2 \omega_k^2 (5 \omega_l^2 + 5 \omega_l \omega_m + 2 \omega_m^2)\right] \nn\\
&&+ \omega_l^2 \omega_m^2 (\omega_l + \omega_m)^2 
+ \omega_k \omega_l \omega_m^2 (\omega_l + \omega_m) (2 \omega_l + \omega_m) 
+ \omega_k^4 (\omega_l^2 + \omega_l \omega_m + \omega_m^2)\nn\\
&&+ \omega_k^3 (2 \omega_l + \omega_m) (\omega_l^2 + \omega_l \omega_m + 2 \omega_m^2)
+ \omega_k^2 (\omega_l^4 + 2 \omega_l^3 \omega_m + 6 \omega_l^2 \omega_m^2 + 5 \omega_l \omega_m^3 + \omega_m^4)\,.
 \ee
 \end{widetext}
%

\bibliographystyle{h-physrev3}
\bibliographystyle{h-physrev3}
\bibliography{bib_source}

\end{document}